\documentclass[11pt]{article}   
\usepackage{geometry}
\usepackage{amsmath,amsfonts,amsthm,enumerate}
\usepackage[dvipsnames]{xcolor}
\usepackage{empheq,tensor,cancel,braket,tcolorbox,blkarray,authblk,cite}
\usepackage{tikz-cd}
\usetikzlibrary{decorations.pathreplacing}
\usepackage[colorlinks=true, linkcolor=blue, citecolor=blue, linktoc=all]{hyperref}
\usepackage{bm}
\usepackage{amssymb}
\usepackage{pifont}
\usepackage{tabularx}
\usepackage{ltablex}
\usepackage{booktabs}
\usepackage{subfigure}

\newcommand{\beq}{\begin{equation}}
\newcommand{\eeq}{\end{equation}}
\newcommand{\td}{\text{d}}
\newcommand{\I}{\text{i}}
\newcommand{\E}{\text{e}}
\newcommand{\nn}{\nonumber}
\newcommand{\p}{\partial}

\allowdisplaybreaks[1]

\title{\huge Entanglement Entropy and Thermodynamics\\
of Dynamical Black Holes}

\author[1]{Weizhen Jia\thanks{weizhenjia@cuhk.edu.hk}}
\author[2]{Qiongyu Qi\thanks{12110204@mail.sustech.edu.cn}}
\author[2]{Christina Gao\thanks{gaoy3@sustech.edu.cn}}

\affil[1]{Department of Physics, The Chinese University of Hong Kong,\protect\\Shatin, New Territories, Hong Kong, China}
\affil[2]{Department of Physics, Southern University of Science and Technology,\protect\\Shenzhen, Guangdong 518055, China}

\date{}							

\begin{document}
\maketitle

\begin{abstract}
We explore the thermodynamic and entanglement properties of dynamical black holes based on the recently proposed dynamical black hole entropy by Hollands-Wald-Zhang. We first provide direct proof that, under first-order perturbations, the dynamical black hole entropy in any $f(R)$ theory equals the Wald entropy evaluated on the generalized apparent horizon. Then, we compute the gravitational entropy explicitly from the replica method using both the event horizon and the apparent horizon as the entangling surfaces, and we show that only the apparent horizon prescription reproduces the correct dynamical black hole entropy satisfying the physical process first law. Furthermore, we reinterpret the generalized second law by identifying the modified von Neumann entropy as the matter entanglement across the apparent horizon. This allows us to express the total entropy as the renormalized generalized entropy evaluated on this surface at the level of the leading local area-law term.
\end{abstract}

\newpage
\begingroup
\hypersetup{linkcolor=black}
\tableofcontents
\endgroup

\newpage
\section{Introduction}

The study of black hole thermodynamics represents a profound synthesis of general relativity, thermodynamics, and quantum physics. This synthesis finds its origins in the pioneering work of Bekenstein \cite{Bekenstein:1972tm,PhysRevD.7.2333}, who postulated that black holes must carry an entropy proportional to their horizon area. This radical proposal emerged from the need to preserve the second law of thermodynamics in the presence of black holes, as the disappearance of matter across an event horizon would otherwise suggest a net decrease in the total entropy of the universe. The subsequent work of Hawking demonstrated that black holes emit thermal radiation at a temperature proportional to their surface gravity \cite{Hawking:1975vcx}, which established the Bekenstein-Hawking entropy formula, $S_{\text{BH}} =\frac{A}{4G}$, as a cornerstone of theoretical physics. This result transformed a thermodynamic analogy into physical reality, which justified that black holes obey the laws of thermodynamics as suggested in \cite{Bardeen:1973gs}.

In the following decades, significant advances were made in understanding black hole entropy within classical gravitational theory beyond general relativity. Derived using the method of covariant phase space, the Wald entropy formula \cite{Wald:1993nt,Iyer:1994ys} interprets black hole entropy as the Noether charge associated with the horizon's Killing symmetry, which provides a general definition of entropy for stationary black holes in any diffeomorphism-invariant theory of gravity:
\begin{align} 
\label{SWald}
S_{\rm Wald}=-2\pi\int_{\cal C}\frac{\delta L}{\delta R_{abcd}}\varepsilon_{ab}\varepsilon_{cd}\,,
\end{align}
where $\cal C$ is a cross section of the future event horizon, and $\varepsilon_{ab}$ is the binormal 2-form to $\cal C$. The Wald entropy represents a natural extension of the Bekenstein-Hawking entropy to theories such as $f(R)$ gravity and Lovelock theories, maintaining the connection between horizon geometry and thermodynamic behavior. This formalism also successfully generalizes the first law of black hole thermodynamics to more general gravitational theories. In their original work \cite{Iyer:1994ys}, Iyer and Wald also proposed a formula for dynamical black hole entropy. However, shortly they noted that this formula is not field redefinition invariant and does not satisfy the second law. Nevertheless, the Wald formalism established a robust framework for equilibrium black hole thermodynamics.

In a parallel development, the concept of entanglement entropy (von Neumann entropy) emerged as a fundamental quantity in quantum field theory and quantum information theory (see, e.g., reviews \cite{Witten:2018zva,Witten:2018zxz}). For a quantum system divided into two subsystems by an entangling surface, the entanglement entropy quantifies the quantum correlations between these subsystems. Specifically, for a spatial region A with complement B, the entanglement entropy is defined as $S_{\text{ent}} = -\text{tr}(\rho_{\rm A} \log \rho_{\rm A})$, where $\rho_{\rm A}$ is the reduced density matrix obtained by tracing over the degrees of freedom in region B. This entropy provides a measure of the quantum entanglement between the two regions and has profound implications for our understanding of quantum many body systems.

The computation of entanglement entropy in quantum field theory can be facilitated by the replica trick \cite{Callan:1994py,Holzhey:1994we,Calabrese:2004eu}, which is a powerful geometric method that involves evaluating the path integral on an $n$-fold Riemann surface branched over the entangling surface. The outcome reveals that the entanglement entropy contains ultraviolet divergences, and the leading divergence scales with the area of the entangling surface. Remarkably, this area law behavior holds even in gravitational contexts, where the entanglement entropy can be computed using similar geometric methods \cite{Lewkowycz:2013nqa}. The holographic principle \cite{tHooft:1993dmi,Susskind:1994vu}, particularly through the AdS/CFT correspondence \cite{Maldacena:1997re, Witten:1998qj}, has provided deep insights into this connection, suggesting that gravitational entropy may have its origins in quantum entanglement \cite{Ryu:2006bv,Hubeny:2007xt,Nishioka:2009un}.

A compelling merger of these developments emerged through the proposal that black hole entropy might be identified as an entanglement entropy. This perspective, initially suggested in \cite{Sorkin:1984kjy} and was further elaborated in \cite{Bombelli:1986rw, Srednicki:1993im,Frolov:1993ym}, posits that the Bekenstein-Hawking entropy originates from the quantum entanglement between degrees of freedom inside and outside the black hole horizon. For an external observer, the states inside the black hole are inaccessible, leading to a mixed-state density matrix for the exterior region. The associated von Neumann entropy naturally exhibits an area law, with the ultraviolet divergence potentially absorbed through the renormalization of the gravitational constant $G$ \cite{Susskind:1994sm,Jacobson:1994iw}. This interpretation has been supported by numerous calculations in both quantum field theory and gravitational physics, though subtleties and disagreements remain unresolved in certain contexts \cite{Callan:1994py, Solodukhin:1994yz, Fursaev:1995ef, Fursaev:1994ea, Demers:1995dq, Kabat:1995eq, Larsen:1995ax, Fursaev:1996uz}.

The thermodynamic interpretation of black hole entropy naturally leads to the formulation of the generalized second law (GSL), which states that the generalized entropy defined as the sum of black hole entropy and the entropy of the matter outside the black hole never decreases. Although it was originally proposed to resolve the violation of the ordinary second law for the matter falls into a black hole \cite{PhysRevD.7.2333}, the GSL found its proper formulation after Hawking's discovery of black hole radiation and the establishment of a consistent thermodynamic interpretation of black holes. Substantial evidence supports the validity of the GSL, particularly through gedanken experiments designed to violate it, which consistently fail in ways analogous to attempts to violate the ordinary second law \cite{Wald:1979zz,PhysRevD.25.942,Unruh:1983ir}.

By taking the von Neumann entropy of the matter outside the black hole as the matter field entropy, a quantum formulation of the GSL in semiclassical general relativity was established in \cite{Wall:2011hj}. As we mentioned, combining the Bekenstein-Hawking entropy and the von Neumann entropy of the matter field yields a finite total entropy with a renormalized gravitational constant $G_{\rm ren}$. Therefore, the generalized entropy can be interpreted as the renormalized Bekenstein-Hawking entropy (ignoring the subleading terms) \cite{Susskind:1994sm}:
\begin{equation}
S_{\text{gen}}[\mathcal{C}(v)] = \frac{1}{4G} A[\mathcal{C}(v)] + S_{\text{vN}}= \frac{1}{4G_{\rm ren}} A[\mathcal{C}(v)]\,.
\end{equation}
For more general theories such as higher curvature theories, the Bekenstein-Hawking entropy is replaced by the Wald entropy $S_{\rm Wald}$, and the subleading divergences in the matter field von Neumann entropy correspond to the renormalization of the higher curvature gravitational couplings appearing in $S_{\rm Wald}$ \cite{Cooperman:2013iqr}.

The extension of black hole thermodynamics to dynamical situations has presented considerable challenges. Two decades after Iyer-Wald attempted to generalize the Noether charge formalism to dynamical black holes, an entropy formula for higher curvature theories was derived by Wall in \cite{Wall:2015raa}. The Wall entropy $S_{\rm Wall}$ includes a correction beyond the Wald entropy that depends on the extrinsic curvature of the horizon, and is valid when a linear non-stationary perturbation is applied to the Killing horizon. However, under the perturbation caused by a flux of ingoing matter, $S_{\rm Wall}$ does not satisfy the ``physical process version'' of the first law \cite{Wald:1995yp, Gao:2001ut} between two arbitrary sections of the future horizon.

Substantial progress in addressing the limitations of previous approaches to dynamical black hole entropy was made in \cite{Hollands:2024vbe,Visser:2024pwz}. By applying the improve Noether charge method, a new notion of dynamical black hole entropy under first-order non-stationary perturbation, denoted by $S_{\text{dyn}}$, was introduced:
\begin{equation}
S_{\text{dyn}}=\frac{2\pi}{\kappa}\int_{\mathcal{C}}\tilde{\mathbf{Q}}_{\xi}=\frac{2\pi}{\kappa}\int_{\mathcal{C}}(\mathbf{Q}_{\xi}-\xi\cdot\mathbf{B}_{\mathcal{H}^{+}})\,,
\end{equation}
where ${\mathbf{Q}}_{\xi}$ is the Noether charge of the horizon Killing field $\xi^a$, and $\mathbf{B}_{\mathcal{H}^{+}}$ is a specific $(n-1)$-form constructed on the future horizon $\mathcal{H}^{+}$. This entropy formula applies to arbitrary diffeomorphism-invariant theories of gravity and satisfies the physical process version of the first law. In Einstein's theory, this entropy takes the form $S_{\text{dyn}} = (1 - v\partial_v)S_{\text{BH}}$,\footnote{This formula for Einstein's theory was also provided in \cite{Rignon-Bret:2023fjq} along with an alternative formula which vanishes on any cross section of a light cone in Minkowski spacetime.} where $v$ is an affine parameter along the horizon. It was also shown that, to the leading perturbative order, $S_{\text{dyn}}$ is equivalent to the Bekenstein-Hawking entropy $S_{\text{BH}}$ of the apparent horizon. In higher curvature theories, the dynamical entropy is shown to be $S_{\text{dyn}} = (1 - v\partial_v)S_{\text{Wall}}$. In particular, since $S_{\rm Wall}$ reduces to $S_{\rm Wald}$ in $f(R)$ gravity, it has been shown in \cite{Kong:2024sqc} that $S_{\text{dyn}}$ for $f(R)$ gravity equals $S_{\rm Wald}$ of the generalized apparent horizon. In Section~\ref{Dynamical Black Hole Entropy from Raychaudhuri Equation}, we will provide an alternative derivation of this based on the discussion for Einstein theory.

Despite these significant advances in understanding the thermodynamic entropy of dynamical black holes, 
its entanglement nature has not been thoroughly investigated.\footnote{In \cite{Dong:2013qoa}, an entanglement entropy for higher derivative gravity has been derived, which agrees with $S_{\rm Wall}$ in all the known cases and is valid under the first-order perturbation. However, it does not satisfy the physical process first law between two arbitrary sections and receives a correction in $S_{\rm dyn}$.} For stationary black holes the event horizon is the natural entangling surface, but its teleological nature calls this choice into question in a dynamical geometry.  Does a locally defined horizon---such as the apparent horizon---take over as the suitable entangling surface under non-stationary perturbations? We will answer this by carrying out replica trick computations on both surfaces. In this paper, we address this fundamental question by investigating the entropy for dynamical black holes in the Einstein gravity and $f(R)$ gravity using the replica method, thereby providing a stepping stone to arbitrary higher curvature theories. Taking the convenience of Gaussian null coordinates (GNC), we compute the entropy associated with both the future event horizon and the apparent horizon as the entangling surface. Our calculations reveal that under first-order perturbations, only the latter agrees with $S_{\text{dyn}}$ and satisfies the physical process first law.
Furthermore, we show that choosing the (generalized) apparent horizon as the entangling surface provides a natural physical interpretation of the modified von Neumann entropy introduced in \cite{Hollands:2024vbe} and further justifies the validity of the GSL for $S_{\text{dyn}}$.

The rest of this paper is organized as follows. In Section~\ref{Dynamical Black Hole Entropy from Raychaudhuri Equation}, we establish the geometric setup for this paper and review the relevant results on dynamical black hole entropy from \cite{Hollands:2024vbe, Visser:2024pwz}. After deriving the expression for $S_{\text{dyn}}$ in Einstein gravity and verifying its equivalence to $S_{\text{BH}}$ of the apparent horizon, we present a direct proof demonstrating that $S_{\text{dyn}}$ in $f(R)$ gravity equals $S_{\text{Wald}}$ evaluated on the generalized apparent horizon. In Section~\ref{Gravitational Entanglement Entropy Around Different Surfaces}, we compute the gravitational entanglement entropy for a dynamical black hole. We take both the future event horizon and the apparent horizon as entangling surfaces and find that the apparent horizon naturally arises as the appropriate boundary for the black hole under non-stationary perturbations. Section~\ref{Modified von Neumann Entropy and the Generalized Second Law} investigates the generalized second law and argues that selecting the apparent horizon as the entangling surface yields a natural physical interpretation of the modified von Neumann entropy. We conclude with a summary and discussion of potential future directions in Section~\ref{Conclusion}. Detailed calculations supporting the results in Section~\ref{Gravitational Entanglement Entropy Around Different Surfaces} are provided in Appendix~\ref{Details of entanglement entropy calculation}. Additionally, a boost weight analysis utilized in the derivations of Section~\ref{Dynamical Black Hole Entropy from Raychaudhuri Equation} can be found in Appendix~\ref{Rescaling Gauge Freedom and Boost Weight}.

\section{Dynamical Black Hole Entropy and the Apparent Horizon}
\label{Dynamical Black Hole Entropy from Raychaudhuri Equation}

In this section, we first provide a brief review of dynamical black hole entropy established in \cite{Hollands:2024vbe, Visser:2024pwz}. We will mainly follow the geometric setup in \cite{Visser:2024pwz}. We begin by setting up the non-stationary geometry in the Gaussian null coordinates and introducing the apparent horizon. Using the Raychaudhuri equation, we then derive the dynamical black hole entropy $S_{\text{dyn}}$ and verify that it coincides with the Bekenstein-Hawking entropy of the apparent horizon. Finally, we extend this result to the generalized apparent horizon in $f(R)$ gravity.

\subsection{Perturbed Geometry}
\label{sec: Gaussian Null Coordinates in Affine Parameterisation}

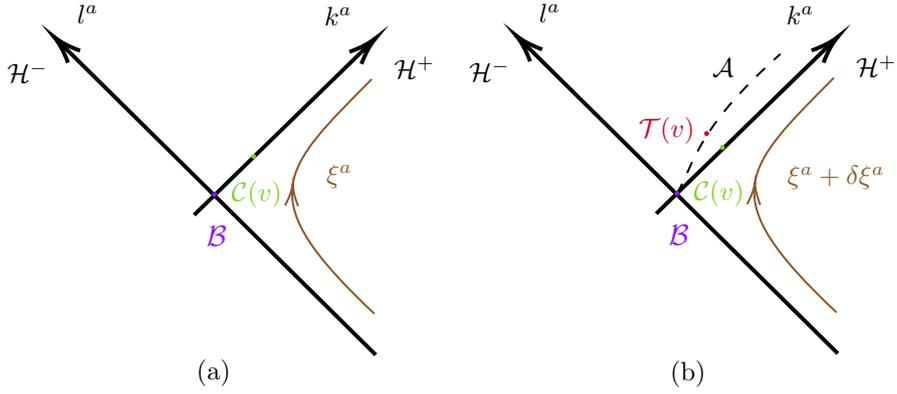
\begin{figure}[h]
    \centering
\label{Bifurcate Killing horizon}

\tikzset{every picture/.style={line width=0.75pt}} 

\begin{tikzpicture}[x=0.75pt,y=0.75pt,yscale=-1,xscale=1]

\draw [line width=1.5]    (240.67,220.67) -- (82.13,62.78) ;
\draw [shift={(80,60.67)}, rotate = 44.88] [color={rgb, 255:red, 0; green, 0; blue, 0 }  ][line width=1.5]    (14.21,-4.28) .. controls (9.04,-1.82) and (4.3,-0.39) .. (0,0) .. controls (4.3,0.39) and (9.04,1.82) .. (14.21,4.28)   ;
\draw [line width=1.5]    (150,150.33) -- (238.53,63.43) ;
\draw [shift={(240.67,61.33)}, rotate = 135.53] [color={rgb, 255:red, 0; green, 0; blue, 0 }  ][line width=1.5]    (14.21,-4.28) .. controls (9.04,-1.82) and (4.3,-0.39) .. (0,0) .. controls (4.3,0.39) and (9.04,1.82) .. (14.21,4.28)   ;
\draw [color={rgb, 255:red, 139; green, 87; blue, 42 }  ,draw opacity=1 ]   (240,200) .. controls (178,141.33) and (193.33,128) .. (238.67,82) ;
\draw [shift={(199.28,134.78)}, rotate = 93.23] [color={rgb, 255:red, 139; green, 87; blue, 42 }  ,draw opacity=1 ][line width=0.75]    (10.93,-3.29) .. controls (6.95,-1.4) and (3.31,-0.3) .. (0,0) .. controls (3.31,0.3) and (6.95,1.4) .. (10.93,3.29)   ;
\draw [line width=1.5]    (471.33,220) -- (312.79,62.12) ;
\draw [shift={(310.67,60)}, rotate = 44.88] [color={rgb, 255:red, 0; green, 0; blue, 0 }  ][line width=1.5]    (14.21,-4.28) .. controls (9.04,-1.82) and (4.3,-0.39) .. (0,0) .. controls (4.3,0.39) and (9.04,1.82) .. (14.21,4.28)   ;
\draw [line width=1.5]    (380.67,149.67) -- (469.19,62.77) ;
\draw [shift={(471.33,60.67)}, rotate = 135.53] [color={rgb, 255:red, 0; green, 0; blue, 0 }  ][line width=1.5]    (14.21,-4.28) .. controls (9.04,-1.82) and (4.3,-0.39) .. (0,0) .. controls (4.3,0.39) and (9.04,1.82) .. (14.21,4.28)   ;
\draw [color={rgb, 255:red, 139; green, 87; blue, 42 }  ,draw opacity=1 ]   (470.67,199.33) .. controls (408.67,140.67) and (424,127.33) .. (469.33,81.33) ;
\draw [shift={(429.94,134.11)}, rotate = 93.23] [color={rgb, 255:red, 139; green, 87; blue, 42 }  ,draw opacity=1 ][line width=0.75]    (10.93,-3.29) .. controls (6.95,-1.4) and (3.31,-0.3) .. (0,0) .. controls (3.31,0.3) and (6.95,1.4) .. (10.93,3.29)   ;
\draw  [dash pattern={on 4.5pt off 4.5pt}]  (391,140) .. controls (400.67,120.67) and (403.33,104.67) .. (442.67,69.33) ;

\draw (221.95,50.33) node  [font=\small]  {$k^{a}$};
\draw (96.99,49.33) node  [font=\small]  {$l^{a}$};
\draw (259.95,76) node  [font=\small]  {$\mathcal{H}^{+}$};
\draw (66.91,78.33) node  [font=\small]  {$\mathcal{H}^{-}$};
\draw (160.33,140.67) node  [font=\LARGE,color={rgb, 255:red, 144; green, 19; blue, 254 }  ,opacity=1 ,rotate=-0.11]  {$\cdot $};
\draw (161.29,160.87) node    {$\mathcal{\textcolor[rgb]{0.56,0.07,1}{B}}$};
\draw (167.33,131.73) node [anchor=north west][inner sep=0.75pt]  [font=\small]  {$\textcolor[rgb]{0.49,0.83,0.13}{\mathcal{C}( v)}$};
\draw (452.61,49.66) node  [font=\small]  {$k^{a}$};
\draw (327.65,48.66) node  [font=\small]  {$l^{a}$};
\draw (490.62,75.33) node  [font=\small]  {$\mathcal{H}^{+}$};
\draw (297.57,77.66) node  [font=\small]  {$\mathcal{H}^{-}$};
\draw (391,140) node  [font=\LARGE,color={rgb, 255:red, 144; green, 19; blue, 254 }  ,opacity=1 ,rotate=-0.11]  {$\cdot $};
\draw (391.96,160.2) node    {$\mathcal{\textcolor[rgb]{0.56,0.07,1}{B}}$};
\draw (413.87,116.97) node  [font=\LARGE,color={rgb, 255:red, 126; green, 211; blue, 33 }  ,opacity=1 ]  {$\cdot $};
\draw (398,131.07) node [anchor=north west][inner sep=0.75pt]  [font=\small]  {$\mathcal{\textcolor[rgb]{0.49,0.83,0.13}{C}}\textcolor[rgb]{0.49,0.83,0.13}{(}\textcolor[rgb]{0.49,0.83,0.13}{v}\textcolor[rgb]{0.49,0.83,0.13}{)}$};
\draw (414.11,75.84) node  [font=\small,rotate=-1.51]  {$\mathcal{A}$};
\draw (386.01,107.53) node  [font=\small]  {$\mathcal{\textcolor[rgb]{0.82,0.01,0.11}{T}}\textcolor[rgb]{0.82,0.01,0.11}{(}\textcolor[rgb]{0.82,0.01,0.11}{v}\textcolor[rgb]{0.82,0.01,0.11}{)}$};
\draw (405.87,109.97) node  [font=\LARGE]  {$\textcolor[rgb]{0.82,0.01,0.11}{\cdot }$};
\draw (179.87,121.64) node  [font=\LARGE,color={rgb, 255:red, 126; green, 211; blue, 33 }  ,opacity=1 ]  {$\cdot $};
\draw (222.62,130.66) node  [font=\small]  {$\textcolor[rgb]{0.55,0.34,0.16}{\xi ^{a}}$};
\draw (470.55,131.57) node  [font=\small]  {$\textcolor[rgb]{0.55,0.34,0.16}{\xi ^{a} +\delta \xi ^{a}}$};
\draw (159.75,229.53) node  [font=\small]  {(a)};
\draw (396.58,230.2) node  [font=\small]  {(b)};

\end{tikzpicture}
\caption{(a) Bifurcate Killing horizon $\mathcal{H}$ consists of the future horizon $\mathcal{H}^+$ and the past horizon $\mathcal{H}^-$, intersecting at the bifurcation surface $\mathcal{B}$. The horizon Killing field is $\xi^{a}$, while $k^a = (\partial_v)^a$ and $l^a = (\partial_u)^a$ are future-directed tangent vectors to the affinely parametrized null geodesics of $\mathcal{H}^+$ and $\mathcal{H}^-$, respectively. $\mathcal{C}(v)$ is a cross section of $\mathcal{H}^+$ at an affine time $v$. (b) The apparent horizon $\cal A$ deviates from ${\cal H}^+$ due to the perturbation. $\mathcal{T}(v)$ is a cross section of $\cal A$ at an affine time $v$.}
\label{Figure 1}
\end{figure}

We consider a dynamical black hole geometry perturbed from a stationary black hole background geometry $(\mathcal{M},g)$ in $d$-dimensional spacetime. The event horizon of the stationary black hole is taken to be a bifurcate Killing horizon $\mathcal{H}$, where we denote the future horizon as $\mathcal{H}^+$, the past horizon as $\mathcal{H}^-$, and the codimension-2 bifurcation surface as $\mathcal{B}$ (see Figure~\ref{Figure 1}). For a stationary black hole, we can define the Killing vector field $\xi^{a}$, which is normal to the Killing horizon. All the background dynamical fields, including the metric $g$ and matter fields will be required to satisfy the Killing symmetry generated by $\xi^{a}$.

One can always establish a null basis $(k^a, l^a)$ on $\mathcal{H}^+$, where $k^a$ represents the future-directed null normal to $\mathcal{H}^+$ and is affinely parametrized as $k^a \overset{\mathcal{H}^+}{=} (\partial_v)^a$, with $v=0$ on $\mathcal{B}$. This prescription does not uniquely determine the affine parameter, since there remains a scaling freedom $v \rightarrow av$, where the constant $a$ may vary from generator to generator but remains fixed along each individual generator of $\mathcal{H}^+$. The vector $l^a$ denotes a future-directed ingoing auxiliary null vector field satisfying $k^a l_a = -1$ on the horizon. We may extend $l^a$ off the horizon by solving the geodesic equation $l^b \nabla_b l^a = 0$ and denote the affine null distance away from $\mathcal{H}^+$ by $u$, thereby identifying $l^a = (\partial_u)^a$. The location of the future horizon is fixed at $u = 0$. Finally, we can extend $k^a$ off the horizon such that it commutes with $l^a$, namely $[k, l]^a = 0$. The metric on $\mathcal{H}^+$ can be decomposed as
\begin{equation}
    g_{ab} \overset{\mathcal{H}^+}{=} -k_a l_b - l_a k_b + \gamma_{ab}\,,
    \label{double null decomposition}
\end{equation}
where $\gamma_{ab}$ is the intrinsic codimension-2 spatial metric of the cross section $\mathcal{C}(v)$ satisfying the orthogonality conditions $\gamma_{ab} k^a \overset{\mathcal{H}^+}{=} \gamma_{ab} l^a \overset{\mathcal{H}^+}{=} 0$. The coordinates $x^i$ can be defined via the projection operator $\gamma^a{}_b \overset{\mathcal{H}^+}{=} \delta^a{}_b + k^a l_b + l^a k_b$, which eliminates components along $k^a$ and $l^a$. The coordinates $(v, u, x^i)$ are called the \textit{Gaussian null coordinates} (GNC). In this coordinate system, we can introduce the following basis vectors:
\begin{equation}
\label{GNCbasis}
k^a = \left( \frac{\partial}{\partial v} \right)^a\, , \qquad l^a = \left( \frac{\partial}{\partial u} \right)^a \,, \qquad m_i^a = \left( \frac{\partial}{\partial x^i} \right)^a\,,
\end{equation}
and the line element of the metric $g_{ab}$ can be written in the following form:  
\begin{equation}
\td s^2 = -2\td u \td v - u^2 \alpha (v, u, x^i) \td v^2 - 2u \omega_i (v, u, x^i) \td v \td x^i + \gamma_{ij} (v, u, x^i) \td x^i \td x^j\,,
\label{2.7}
\end{equation}
with the corresponding inverse metric  
\begin{equation}
g^{ab} = 
\begin{pmatrix}
u^2 (\alpha + \omega_i \omega^i) & -1 & -u \omega^i \\
-1 & 0 & 0 \\
-u \omega^i & 0 & \gamma^{ij}
\end{pmatrix}\,,
\label{inverse metric}
\end{equation}
where $\gamma^{ij}$ denotes the inverse of $\gamma_{ij}$, and $\omega^i = \gamma^{ij} \omega_j$. In the GNC gauge, the metric components $g_{uu}$ and $g_{ui}$ vanish throughout the spacetime. The form of $g_{vv}$ is determined by the affine geodesic equation of $k^{a}$ on the horizon, where the function $\alpha$ is non-singular at $u=0$ or $v=0$ \cite{Visser:2024pwz}. Furthermore, the fact that $g_{vi}$ vanishes on ${\cal H}^+$ and $g_{ij}$ remains regular there implies that $\omega_i$ and $\gamma_{ij}$ are also non-singular at $u=0$ or $v=0$. The basis vectors in \eqref{GNCbasis} have the following dual vectors:
\begin{equation}
\begin{split}
k_a &= -(\td u)_a - u^2 \alpha (\td v)_a - u \omega_i (\td x^i)_a\,,\ \\l_a &= -(\td v)_a\,, \qquad m_a^i = (\td x^i)_a - u\omega^i(\td v)_a \,.
\end{split}
\label{2.12}
\end{equation}
It is easy to verify that the metric indeed satisfies the double null decomposition on the horizon, namely $g_{ab} \overset{\mathcal{H}^+}{=} -2k_{(a} l_{b)} + \gamma_{ab}$, where the transverse metric can be expressed as
\begin{equation}
\gamma_{ab} = \gamma_{ij}(\td x^i)_a(\td x^j)_b \overset{\mathcal{H}^+}{=} \gamma_{ij}m_a^i m_b^j\,.
\label{2.13}
\end{equation}

At the future horizon, the Killing vector $\xi^a$ satisfies $\xi^b \nabla_b \xi^a \overset{\mathcal{H}^+}{=} \kappa \xi^a$, and the null normal satisfies $k^b \nabla_b k^a \overset{\mathcal{H}^+}{=} 0$. On the other hand, at the past horizon we have $\xi^{a}\nabla_{a}\xi^{b}\overset{\mathcal{H}^{-}}{=}-\kappa\xi^{b}$ and $l^{a}\nabla_{a}l^{b}=0$. Hence, we can rescale $k^a$ and $l^{a}$ such that
\begin{equation}
\label{Killing}
    \xi^a  = \kappa(vk^{a}-ul^{a})\,,
\end{equation}
where $\kappa$ is the surface gravity. Therefore, the metric functions for the unperturbed stationary geometry, denoted by $\alpha^{(0)}$, $\omega^{(0)}_i$, and $\gamma^{(0)}_{ij}$, are subject to additional constraints. Specifically, the isometry generated by the Killing vector $\xi^a$ restricts their form so that the dependence on $u$ and $v$ occurs only through the combination $\kappa u v$ \cite{Hollands:2022fkn, Bhattacharyya:2021jhr}:
\begin{equation}
    \alpha^{(0)}=\alpha^{(0)}(\kappa uv,x^{i})\,,\qquad\omega^{(0)}_i=\omega^{(0)}_i(\kappa uv,x^{i})\,,\qquad\gamma^{(0)}_{ij}=\gamma^{(0)}_{ij}(\kappa uv,x^{i})\,.
\end{equation}

In order to investigate the non-stationary case, we consider the metric as perturbed from the stationary background $g^{(0)}_{ab}$ as $g_{ab}=g^{(0)}_{ab}+\delta g_{ab}$ and the matter fields as $\phi=\phi^{(0)}+\delta \phi$, where $\delta g_{ab}, \delta \phi \sim \mathcal{O}(\epsilon)$ are first-order perturbations and $\epsilon\ll 1 $ is a small perturbation parameter. Of course, once we turn on a non‐stationary perturbation the exact Killing symmetry is lost and, in principle, the location and parametrization of the horizon could shift in complicated ways. However, since we are working only to the first order around a stationary background, we may exploit the diffeomorphism gauge freedom to keep the identification of points between the two spacetimes simple, and to preserve the affine parametrization of the null generators. This puts the perturbed geometry under control and allows us to track the horizon perturbatively. Concretely, we follow \cite{Visser:2024pwz} and impose the following gauge conditions:
\begin{enumerate}[(1)]
    \label{gauge condition 1}
    \item The event horizon of the perturbed black hole is identified with the Killing horizon of the unperturbed background. Moreover, $\mathcal{H}^+$ and $\mathcal{H}^-$ remains at $u = 0$ and $v = 0$, respectively, after the perturbation is applied.
    \label{Condition 2}
    \item The null vectors $k^a$ and $l^a$ are held fixed under the perturbation
    \begin{equation}
        \delta k^a = 0\,, \qquad \delta l^a = 0\,.
    \end{equation}
    Additionally, $k^a$ remains null and normal to $\mathcal{H}^+$, while $l^a$ stays null everywhere under the perturbation. Combined with the requirement $\delta(k^a l_a) = 0$, this implies that
    \begin{equation}
        k^a \delta g_{ab} \overset{\mathcal{H}^+}{=} 0\,, \qquad l^a \delta g_{ab} = 0\,.
    \end{equation}
    We further require that $k^a$ remains affinely parametrized on $\mathcal{H}^+$, and $l^a$ remains affinely parametrized in the entire spacetime under the perturbation, i.e.,
    \begin{equation}
        \delta(k^b \nabla_b k^a) \overset{\mathcal{H}^+}{=} 0\,, \qquad \delta(l^b \nabla_b l^a) = 0\,.
    \end{equation}
    \item The perturbed vector field $\xi^a$ remains null and tangent to the geodesic generators of the perturbed black hole. Along with $\xi^a \delta g_{ab} \overset{\mathcal{H}}{=} 0$, this ensures that $\delta \xi^a$ is proportional to $k^a$ on $\mathcal{H}^+$ and to $l^a$ on $\mathcal{H}^-$.
    \label{Condition 3}
\end{enumerate}

From the above gauge conditions, it follows that the affine parameters $v$ and $u$ are fixed under the perturbation. We will also fix the spatial coordinates $x^i$. Therefore, a non-stationary perturbation of a stationary background affects only the metric functions in the following way: 
\begin{align}
\alpha( u,v, x^i) &= \alpha^{(0)}(\kappa uv, x^i) + \delta\alpha(v, u, x^i)\,,
\label{dynamical perturbation}
\\\omega_i( u,v, x^i) &= \omega^{(0)}_i(\kappa uv, x^i) + \delta\omega_i(v, u, x^i)\,,
\\\gamma_{ij}( u,v, x^i) &= \gamma^{(0)}_{ij}(\kappa uv, x^i) + \delta\gamma_{ij}(v, u, x^i)\,.
\end{align}

\subsection{Apparent Horizon}
Now we introduce the apparent horizon in this perturbed geometry. An apparent horizon is defined geometrically as the boundary separating regions with negative outgoing null expansion from those with non-negative expansion. This hypersurface is foliated by marginally outer trapped surfaces where the outgoing null expansion vanishes identically, representing the instantaneous boundary of trapped regions within a given Cauchy slice. For a stationary black hole, the apparent horizon coincides with the event horizon.

Recall that the outgoing null expansion $\theta_{v} = \nabla_{a}k^{a}$ of the future horizon is equal to the rate of change of the area element $\mathrm{d}A$ along the affine null parameter $v$:
\begin{equation}
\theta_{v}\,\mathrm{d}A = \frac{\mathrm{d}}{\mathrm{d}v}(\mathrm{d}A)\,,
\label{2.17}
\end{equation}
where $\mathrm{d}A = \sqrt{\det\gamma}\,\mathrm{d}^{d-2}x$ is the area element of a cross section of the horizon. From \eqref{2.17}, it is easy to verify that
\begin{equation}
\label{thetav}
\frac{1}{2} \gamma^{ac} \partial_v \gamma_{ac} = \frac{1}{\sqrt{\det\gamma}} \partial_v \sqrt{\det\gamma} = \theta_v\,. 
\end{equation}
In the GNC system, the position of the apparent horizon $\mathcal{A}$ is denoted as $u = \mathcal{U}(v, x^i) \geqslant 0$, which represents the affine null distance from the event horizon, which is nonzero under perturbation, as shown in Figure~\ref{Figure 1}~(b). Throughout our analysis, we assume that $\mathcal{U}(v, x^i)$ and its spacetime derivatives are of the first order in the perturbation parameter $\epsilon$. 

A cross section $\mathcal{T}(v)$ of the apparent horizon $\mathcal{A}$ located at an affine time $v$ is a future marginally outer trapped surface. Later on, we will also refer to $\cal T$ as the apparent horizon for short when there is no confusion. One of the null normal covectors of $\mathcal{T}$ is given by $l_a = -(\td v)_a$. Through the constraint $\td u - (\partial_i \mathcal{U}) \td x^i \overset{\mathcal{T}}{\rightarrow} 0$, the other null normal covector of $\mathcal{T}$ can be determined to be
\begin{equation}
\tilde{k}_a \overset{\mathcal{T}}{=} -(\td u)_a + \partial_i \mathcal{U}(\td x^i)_a\,.
\label{2.26}
\end{equation}
Rasing the index using \eqref{inverse metric}, we have
\begin{equation}
\label{tildeka}
\tilde{k}^a = g^{ab} \tilde{k}_b \overset{\mathcal{T}}{=}   k^a + (D^i \mathcal{U} + \omega^i \mathcal{U}) m_i^a + \mathcal{O}(\epsilon^2)\,,
\end{equation}
where $D_i$ represents the codimension-2 intrinsic covariant derivative, and $D^i = \gamma^{ij} D_j$. 

It is easy to verify that $l^{a}$ and $\tilde{k}^{a}$ satisfy the consistency requirements $\tilde{k}^a \tilde{k}_a = \mathcal{O}(\epsilon^2)$ and $\tilde{k}^{a}l_{a}=-1$. By means of these two null normal vectors, the metric has the following decomposition on ${\cal T}$, which is valid to the first order in $\epsilon$:
\begin{equation}
g_{ab}\stackrel{{\cal T}}{{=}}-2\tilde{k}_{(a}l_{b)}+\gamma_{ab}\,. 
\label{metric decomposition}
\end{equation}
Then, the expansion of the apparent horizon $\mathcal{T}$ along the direction of $\tilde{k}^a$ is
\begin{equation}
\tilde{\theta}_{\tilde{k}}=\gamma^{a}{}_{b}\nabla_{a}\tilde{k}^{b}\stackrel{{\mathcal{T}}}{{=}}\gamma^{a}{}_{b}\nabla_{a}k^{b}+\gamma^{a}{}_{b}\nabla_{a}((D^{i}{\cal U}+\omega^{i}{\cal U})m_{i}^{b})=\theta_{v}+D_{i}(D^{i}{\cal U}+\omega^{i}{\cal U})+{\cal O}(\epsilon^{2})\,. 
\label{2.30}
\end{equation}
where we used the orthogonality relation $m_{a}^{j}m_{i}^{a}=\delta_{i}^{j}$ and the fact that $\gamma^{a}{}_{b}$ is a projection operator to the first order. Note that the characteristic condition for an apparent horizon is that the expansion $\tilde{\theta}_k$ along $\tilde{k}$ must vanish, while $\theta_v$ needs not vanish on $\mathcal{T}$ since $k^a$ is not a normal of $\mathcal{T}$. Then, on the apparent horizon we have
\begin{equation}
\tilde{\theta}_{\tilde k} \equiv \theta_v (v, \mathcal{U}, x^i) + D_i (D^i \mathcal{U} + \omega^i \mathcal{U})\overset{\mathcal{T}}{=}0\,. 
\label{the characteristic condition}
\end{equation}
Expanding \eqref{the characteristic condition} around the event horizon $u=0$, to the first-order perturbation we have
\begin{align}
0 &\overset{\mathcal{T}}{=} \theta_v (v, 0, x^i) + \mathcal{U}(v, x^i) \partial_u \theta_v (v, 0, x^i) + D_i (D^i \mathcal{U} + \omega^i \mathcal{U})\nn\\
&=\theta_v (v, 0, x^i) + \mathcal{U}(v, x^i) F^\theta_1(x^{i}) + D_i (D^i \mathcal{U} + \omega^i \mathcal{U})\,,
\label{2.36}
\end{align}
where we used the fact that $\theta_{u}(v, 0, x^i)=vF^\theta_1(x^{i})$ with $F^\theta_1(x^{i})$ being a negative function at the zeroth order (see Appendix~\ref{Rescaling Gauge Freedom and Boost Weight}), and we used the relation $\partial_{u}\theta_{v}=\partial_{v}\theta_{u}$ of two null expansions at the zeroth order. Note that $\theta_v (v, 0, x^i)$ vanishes before turning on the perturbation, and hence it can also be interpreted as $\delta\theta_v (v, 0, x^i)$. In principle, by solving \eqref{2.36}, one can find $\mathcal{U}(v, x^i)$ explicitly and determine the location of the apparent horizon. In the next subsection, we will use the above geometry setup to derive the dynamical black hole entropy and verify the relation with the Bekenstein-Hawking entropy of the apparent horizon.

\subsection{Dynamical Black Hole Entropy for Einstein's Gravity}

First, we review the derivation of the ``physical process'' version of the first law for non-stationary perturbations of a stationary black hole following \cite{Visser:2024pwz}, which is based on the Raychaudhuri equation and applies to Einstein's gravity. Then we will examine the relationship between the dynamical black hole and the area of the apparent horizon.

\begin{figure}[h]
\centering

\tikzset{every picture/.style={line width=0.75pt}} 

\begin{tikzpicture}[x=0.75pt,y=0.75pt,yscale=-1,xscale=1]

\draw [line width=1.5]    (282.67,228.67) -- (124.13,70.78) ;
\draw [shift={(122,68.67)}, rotate = 44.88] [color={rgb, 255:red, 0; green, 0; blue, 0 }  ][line width=1.5]    (14.21,-4.28) .. controls (9.04,-1.82) and (4.3,-0.39) .. (0,0) .. controls (4.3,0.39) and (9.04,1.82) .. (14.21,4.28)   ;
\draw [line width=1.5]    (192,158.33) -- (280.53,71.43) ;
\draw [shift={(282.67,69.33)}, rotate = 135.53] [color={rgb, 255:red, 0; green, 0; blue, 0 }  ][line width=1.5]    (14.21,-4.28) .. controls (9.04,-1.82) and (4.3,-0.39) .. (0,0) .. controls (4.3,0.39) and (9.04,1.82) .. (14.21,4.28)   ;
\draw  [dash pattern={on 4.5pt off 4.5pt}]  (202.33,148.67) .. controls (212,129.33) and (214.67,113.33) .. (254,78) ;
\draw    (247,124.33) -- (229.02,104.8) ;
\draw [shift={(227.67,103.33)}, rotate = 47.37] [color={rgb, 255:red, 0; green, 0; blue, 0 }  ][line width=0.75]    (10.93,-3.29) .. controls (6.95,-1.4) and (3.31,-0.3) .. (0,0) .. controls (3.31,0.3) and (6.95,1.4) .. (10.93,3.29)   ;

\draw (263.95,58.33) node  [font=\small]  {$k^{a}$};
\draw (138.99,57.33) node  [font=\small]  {$l^{a}$};
\draw (301.95,84) node  [font=\small]  {$\mathcal{H}^{+}$};
\draw (108.91,86.33) node  [font=\small]  {$\mathcal{H}^{-}$};
\draw (202.33,148.67) node  [font=\LARGE,color={rgb, 255:red, 144; green, 19; blue, 254 }  ,opacity=1 ,rotate=-0.11]  {$\cdot $};
\draw (203.29,168.87) node    {$\mathcal{\textcolor[rgb]{0.56,0.07,1}{B}}$};
\draw (225.21,125.64) node  [font=\LARGE,color={rgb, 255:red, 126; green, 211; blue, 33 }  ,opacity=1 ]  {$\cdot $};
\draw (209.33,139.73) node [anchor=north west][inner sep=0.75pt]  [font=\small]  {$\mathcal{\textcolor[rgb]{0.49,0.83,0.13}{C}}\textcolor[rgb]{0.49,0.83,0.13}{(}\textcolor[rgb]{0.49,0.83,0.13}{v}\textcolor[rgb]{0.49,0.83,0.13}{_{1})}$};
\draw (236.78,72.51) node  [font=\small,rotate=-1.51]  {$\mathcal{A}$};
\draw (197.34,116.2) node  [font=\small]  {$\mathcal{\textcolor[rgb]{0.82,0.01,0.11}{T}}\textcolor[rgb]{0.82,0.01,0.11}{(}\textcolor[rgb]{0.82,0.01,0.11}{v}\textcolor[rgb]{0.82,0.01,0.11}{_{1})}$};
\draw (217.21,117.97) node  [font=\LARGE]  {$\textcolor[rgb]{0.82,0.01,0.11}{\cdot }$};
\draw (251.21,100.31) node  [font=\LARGE,color={rgb, 255:red, 126; green, 211; blue, 33 }  ,opacity=1 ]  {$\cdot $};
\draw (239.21,91.97) node  [font=\LARGE]  {$\textcolor[rgb]{0.82,0.01,0.11}{\cdot }$};
\draw (212.67,89.53) node  [font=\small]  {$\mathcal{\textcolor[rgb]{0.82,0.01,0.11}{T}}\textcolor[rgb]{0.82,0.01,0.11}{(}\textcolor[rgb]{0.82,0.01,0.11}{v}\textcolor[rgb]{0.82,0.01,0.11}{_{2}}\textcolor[rgb]{0.82,0.01,0.11}{)}$};
\draw (256.67,102.4) node [anchor=north west][inner sep=0.75pt]  [font=\small]  {$\mathcal{\textcolor[rgb]{0.49,0.83,0.13}{C}}\textcolor[rgb]{0.49,0.83,0.13}{(}\textcolor[rgb]{0.49,0.83,0.13}{v}\textcolor[rgb]{0.49,0.83,0.13}{_{2}}\textcolor[rgb]{0.49,0.83,0.13}{)}$};
\draw (261.23,132.85) node  [font=\small]  {$\delta T_{ab}$};

\end{tikzpicture}
\caption{The apparent horizon deviates from the event horizon due to the non-stationary perturbation. $\delta T_{ab}$ represents the infalling matter field that causes the perturbation near the future event horizon $\mathcal{H}^+$. The entropy change between two generic cross sections $\mathcal{C}(v_2)$ and $\mathcal{C}(v_1)$ is related to the matter Killing energy flux through the physical process first law.}
\label{Fig2}

\end{figure}
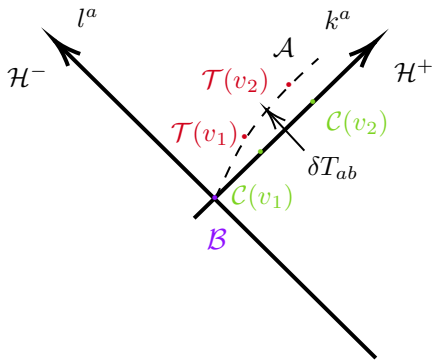

We perturb the stationary geometry by introducing matter with stress-energy tensor $\delta T_{ab}$ that crosses the horizon \cite{Hawking:1972hy, Hollands:2024vbe, Visser:2024pwz}, assuming the perturbation is sufficiently small to preserve the event horizon (see Figure~\ref{Fig2}). The standard physical process first law \cite{Wald:1995yp, Gao:2001ut, Poisson:2009pwt} relates entropy changes to the absorbed mass and angular momentum of this infalling matter, but requires stationary initial/final states where the horizon coincides with the Killing horizon at $v=0$ and $v=+\infty$. By relaxing these assumptions and allowing non-stationary states at arbitrary affine times $v_1$ and $v_2$, we will see that the Bekenstein-Hawking entropy receives a dynamical correction, which leads to the area law of the apparent horizon.

Suppose the horizon Killing vector is normalized as $\xi^{a} = (\partial_{t})^{a} + \Omega_{\mathcal{H}} (\partial_\varphi)^{a}$, where $\Omega_{\mathcal{H}}$ denotes the angular velocity of the horizon, then the matter Killing energy flux through the future horizon between two arbitrary cross sections ${\cal C}(v_{1})$ and ${\cal C}(v_{2})$ is related to the changes in the mass and angular momentum of the matter crossing the horizon via \cite{Gao:2001ut, Poisson:2009pwt, Hollands:2024vbe, Visser:2024pwz}
\begin{equation}
\int_{v_{1}}^{v_{2}}\text{d}v\int_{\mathcal{C}}\text{d}A\,\delta T_{ab} \xi^{a}k^{b} = \Delta\delta M - \Omega_{\mathcal{H}}\,\Delta\delta J\,,
\label{2.16}
\end{equation}
where $\Delta$ denotes the difference between two horizon cross sections, and $\mathrm{d}A=\sqrt{\det\gamma}\,\mathrm{d}^{d-2}x$ is the area element of a cross section ${\cal C}$ of the horizon at an affine time $v$.

The Raychaudhuri equation for the null geodesic congruence generated by $k^a$ reads
\begin{equation}
\frac{\mathrm{d}\theta_{v}}{\mathrm{d}v} = -\frac{1}{d-2}\theta_{v}^{2} - \sigma^{ab}\sigma_{ab} + \omega^{ab}\omega_{ab} - R_{ab}k^{a}k^{b}\,,
\label{2.18}
\end{equation}
where $\sigma_{ab}$ and $\omega_{ab}$ are the shear and twist of the null congruence. Notice that $\omega_{ab} \overset{\mathcal{H}^+}{=} 0$ since $k^{a}$ is orthogonal to $\mathcal{H}^+$, and $\theta_{v}, \sigma_{ab}$ are both first-order quantities in the perturbation. Furthermore, in the linear approximation, the perturbation of the Einstein equation satisfies $\delta R_{ab}k^{a}k^{b} = 8\pi G \delta T_{ab}k^{a}k^{b}$. Thus, the perturbation of the Raychaudhuri equation to the first order gives
\begin{equation}
\frac{\mathrm{d}\delta\theta_{v}}{\mathrm{d}v} =- 8\pi G \delta T_{ab}k^{a}k^{b}\,.
\label{2.18}
\end{equation}
Note that this also holds if one applies Einstein's equation with a cosmological constant.

From the gauge condition \eqref{Condition 2}, on the future horizon we have $\xi^{a} \stackrel{\mathcal{H}^+}{=} \kappa v k^{a}$. Multiplying the varied Raychaudhuri equation by $\kappa v$ and integrating over the horizon between the affine parameters $v_1$ and $v_2$, we obtain
\begin{equation}
\label{perturbedRaychaudhuri}
\kappa \int_{v_1}^{v_2} \mathrm{d}v \int_{\mathcal{C}} \mathrm{d}A  v \frac{\mathrm{d} \delta \theta_v}{\mathrm{d} v} = -8\pi G \int_{v_1}^{v_2} \mathrm{d}v \int_{\mathcal{C}} \mathrm{d}A  \delta T_{ab} \xi^{a} k^{b}.
\end{equation}
The left-hand side of the above equation can be evaluated as follows:
\begin{align}
&\int_{\mathcal{C}} \mathrm{d}A \int_{v_1}^{v_2} \mathrm{d}v v \frac{\mathrm{d} \delta \theta_v}{\mathrm{d} v} =  - \int_{v_1}^{v_2} \mathrm{d}v \int_{\mathcal{C}}\mathrm{d}A \delta \theta_v+\int_{\mathcal{C}} \mathrm{d}A \left( v \delta \theta_v \right)\bigg|_{v_1}^{v_2} \nn\\
\label{Sdynderive}
 ={}& - \int_{v_1}^{v_2} \mathrm{d}v \delta\int_{\mathcal{C}}\mathrm{d}A  \theta_v+\delta \int_{\mathcal{C}} \mathrm{d}A \left( v \theta_v \right)\bigg|_{v_1}^{v_2}  = -\Delta \delta \left( \int_{\mathcal{C}} \mathrm{d}A (1 - v \theta_v) \right)\,,
\end{align}
where in the first equality we integrated by parts, in the second equality the variation is taken outside the integral as $\theta_v$ vanishes for the unperturbed geometry on any ${\cal C}$, and in the last step we applied \eqref{2.17}. As we can see, in addition to the area term, there is a nontrivial dynamical correction to the Bekenstein-Hawking entropy, which is absent in the standard derivation \cite{Wald:1995yp} of the physical process first law.

On the other hand, the right-hand side of \eqref{perturbedRaychaudhuri} gives rise to the matter Killing energy flux in \eqref{2.16}. Therefore, the physical process first law applicable to arbitrary cross sections of the future horizon under non-stationary perturbations reads
\begin{equation}
\frac{\kappa}{2\pi} \Delta \delta S_{\text{dyn}} = \Delta \delta M - \Omega_{\mathcal{H}} \Delta \delta J\,, 
\label{first law}
\end{equation}
where $S_{\text{dyn}}$ denotes the dynamical black hole entropy associated with the cross section $\mathcal{C}$ of the horizon, whose expression can be read off from \eqref{Sdynderive} to be
\begin{equation}
S_{\text{dyn}}[\mathcal{C}] = \frac{1}{4G} \int_{\mathcal{C}} \td A (1 - v\theta_v)\,. 
\label{the dynamical black hole entropy of the cross section}
\end{equation}
Under the assumption of the null energy condition $\delta T_{ab}k^a k^b \geqslant 0$, the matter Killing energy flux is always non-negative, which ensures that the dynamical black hole entropy satisfies the second law of black hole thermodynamics in the linear perturbation:
\begin{equation}
\Delta \delta S_{\text{dyn}} \geqslant 0\,. 
\label{``linearised" second law}
\end{equation}

Finally, we demonstrate the relation between $S_{\text{dyn}}$ and the area of the apparent horizon. The Taylor expansion of the area element $\text{d}A$ up to first order gives
\begin{align}
    \td A(v,{\cal U},x^{i})&=\td A (v, 0, x^i) + \mathcal{U}(v, x^i) \partial_u \text{d}A (v, 0, x^i)\nn\\
    &=\td A (v, 0, x^i) + \mathcal{U}(v, x^i) vF^\theta_1(x^{i}) \td A (v, 0, x^i)\nn\\
    &=\text{d}A (v, 0, x^i)-(v\theta_v(v, 0, x^i)+vD_{i}(D^{i}{\cal U}+\omega^{i}{\cal U}))\td A(v, 0, x^i)\nn\\
    \label{deltadAU}
    &=(1-v\partial_{v})\td A(v,0,x^{i})-v{\rm d}A(v,0,x^{i})D_{i}(D^{i}{\cal U}+\omega^{i}{\cal U})\,,
\end{align}
where the relation $\partial_{u}\td A=\theta_{u}\td A=vF^\theta_1(x^{i})\td A$ is used in the second equality, we used \eqref{2.36} in the third equality, and \eqref{2.17} is used in the last step. Noticing that $\td A=\sqrt{\det\gamma}\td^{d-2}x$, the second term on the right-hand side as a total derivative will not contribute after integrating over the codimension-2 surface ${\cal C}(v)$. Then, after integration we have
\begin{equation}
A(v,{\cal U})=(1-v\partial_{v})A(v,0)\,.
\end{equation}
Therefore, with the correction from the non-stationary perturbation, dynamical black hole entropy in Einstein's gravity corresponds to precisely the area of the apparent horizon, i.e.,
\begin{equation}
S_{\rm dyn}=\frac{1}{4G}(1-v\partial_{v})A[{\cal C}(v)]=\frac{A[ {\cal T}(v)]}{4G}\,.
\label{2.45}
\end{equation}

\subsection{Dynamical Black Hole Entropy for $f(R)$ Gravity}
\label{Relation to Wald Entropy of the Apparent Horizon in}

For an arbitrary theory of gravity with a diffeomorphism covariant Lagrangian, such as $f(R)$ and $f(\text{Riemann})$ theories, the first law of a dynamical black hole is derived in \cite{Hollands:2024vbe, Visser:2024pwz} by applying the improved Noether charge method. In $f(R)$ gravity, the dynamical black hole entropy is given by replacing the Bekenstein-Hawking entropy in Einstein's gravity with the Wald entropy:
\begin{equation}
\label{SdynSwaldfR}
    S_{\text{dyn}}=\frac{1}{4G}(1-v\partial_{v})\int_{\mathcal{C}}f'(R)=(1-v\partial_{v})S_{\rm Wald}[{\cal C}(v)]\,,
\end{equation}
where $f'(R)\equiv \td f(R)/\td R$. The dynamical entropy $S_{\text{dyn}}$ in $f(R)$ gravity includes a correction to the Wald entropy of a stationary black hole, analogous to how $S_{\text{dyn}}$ in general relativity corrects the Bekenstein-Hawking entropy. The above expression for $S_{\text{dyn}}$ was also derived in \cite{Kong:2024sqc} by means of the Einstein frame, where it was verified to satisfy both the physical process first law and the linearized second law. It was also shown in \cite{Kong:2024sqc} that the $S_{\text{dyn}}$ for $f(R)$ gravity corresponds to the Wald entropy of the generalized apparent horizon. Working in GNC, in this subsection we provide an alternate proof of this similar to the proof for Einstein's gravity in the previous subsection.

In $f(R)$ gravity, the \emph{generalized expansion} of the null congruence generated by $k^a=(\p_v)^a$ is defined as follows \cite{Matsuda:2020yvl}:
\begin{equation}
 \label{definition of generalized expansion}
    \Theta_{v}=\frac{\partial}{\partial v}\log(\sqrt{\det\gamma}f'(R))=\theta_v+k^a\nabla_a\log(f'(R))\,.
\end{equation}
For a stationary background, the \emph{generalized apparent horizon} in $f(R)$ gravity is defined as the hypersurface where $\Theta_v = 0$, separating regions of negative and non-negative generalized expansion. Under the non-stationary perturbation, the generalized expansion is perturbed to $\tilde{\cal A}$, whose location is described by $u = \mathcal{U}(v, x^i) \geqslant 0$, which also represents the affine null distance $\mathcal{U}(v, x^i)$ to the event horizon. We assume that $\mathcal{U}(v, x^i)$ and its spacetime derivatives are of the first order in the perturbation. 

Let $\tilde{\mathcal{T}}(v)$ denote the cross section of $\tilde{\mathcal{A}}$ at an affine time $v$. Later on, we will refer to $\tilde{\mathcal{T}}(v)$ as the generalized apparent horizon for short when there is no confusion. As before, the null normals of $\tilde{\mathcal{T}}(v)$ are $l^a$ and $\tilde k^a$ satisfying \eqref{2.26}, \eqref{tildeka} and \eqref{metric decomposition} on $\tilde{\cal T}$. Under the perturbation, we need to consider the generalized expansion of $\tilde{k}^a$:
\begin{equation}
\label{genexpansiontilde}
\tilde{\Theta}_{\tilde{k}}=\tilde{\theta}_{\tilde{k}}+\tilde{k}^{a}\nabla_{a}\text{log}(f'(R))\,.
\end{equation}
Using the geometry settings we have defined before, we can write
\begin{align}
\tilde{\Theta}_{\tilde{k}}&=\gamma^{a}{}_{b}\nabla_{a}\tilde{k}^{b}+\frac{1}{f'(R)}\tilde{k}^{a}\nabla_{a}(f'(R))\nn\\
&\stackrel{{\tilde{\mathcal{T}}}}{{=}}
\Theta_{v}(v, \mathcal{U}, x^i)+D_{i}(D^{i}{\cal U}+\omega^{i}{\cal U})+\frac{1}{f'(R)}(D^{i}\mathcal{U}+\omega^{i}\mathcal{U})D_{i}f'(R)+\mathcal{O}(\epsilon^{2})\,.
\end{align}
The generalized apparent horizon in the perturbed geometry is determined by the vanishing of $\tilde{\Theta}_{\tilde{k}}$, while $\Theta_{v}$ need not vanish on this horizon as $k^a$ is not a normal of $\tilde{\mathcal{T}}$. Then, up to the first order in the perturbation, on the generalized apparent horizon we have
\begin{align}
\label{GAHperturbation}
    0&\overset{\tilde{\mathcal{T}}}{=}\Theta_v (v, 0, x^i) + \mathcal{U}(v, x^i) \partial_u \Theta_v (v, 0, x^i) + D_i (D^i \mathcal{U} + \omega^i \mathcal{U})+\frac{1}{f'(R)}(D^{i}\mathcal{U}+\omega^{i}\mathcal{U})D_{i}f'(R)\,. 
\end{align}
Note that $\Theta_v (v, 0, x^i)$ vanishes before turning on the perturbation, and hence it can also be interpreted as $\delta\Theta_v (v, 0, x^i)$.

To further evaluate the above expression, we will use several properties of the generalized expansion. First, it follows from \eqref{definition of generalized expansion} that
\begin{equation}
\label{pvfRdA}
    \partial_{v}(f'(R)\td A)=\Theta_{v}f'(R)\td A\,.
\end{equation}
Second, to zeroth order in the perturbation we have
\begin{align}
\partial_u \Theta_v &= \partial_u \left ( \frac{1}{f'(R)dA} \partial_v(f'(R) \td A) \right ) \nn\\ 
&= -\frac{1}{(f'(R)\td A)^2} \partial_u (f'(R)\td A)\partial_v(f'(R) \td A) + \frac{1}{f'(R)\td A} \partial_u \partial_v(f'(R) \td A)\nn\\ 
&= \partial_v \left( \frac{1}{f'(R)\td A} \partial_u(f'(R) \td A) \right) = \partial_v \Theta_u\,, 
\label{u expansion}
\end{align}
where $\Theta_u = \partial_{u}\text{log}(f'(R)\td A)$ is the ingoing null generalized expansion in the $u$-direction. Moreover, at the zeroth order $\Theta_u$ satisfies 
\begin{equation}
\Theta_u (v, 0, x^i) = F^\Theta_1(x^i)v\,, 
\label{weight}
\end{equation}
where the subscript $1$ indicates that the boost weight of $\Theta_u$ is $-1$ (see Appendix~\ref{Rescaling Gauge Freedom and Boost Weight}).
Using these facts, we can obtain from \eqref{GAHperturbation} that
\begin{equation}
\Theta_v (v, 0, x^i) = -\mathcal{U}(v, x^i) F^\Theta_1(x^i) - D_i (D^i \mathcal{U} +\omega^i \mathcal{U})-\frac{1}{f'(R)}(D^{i}\mathcal{U}+\omega^{i}\mathcal{U})D_{i}f'(R)\,. 
\label{solution}
\end{equation}

To relate the dynamical black hole entropy \eqref{SdynSwaldfR} in $f(R)$ gravity to the generalized apparent horizon, we consider the Taylor expansion of $f'(R)\td A$ to the first order. Following a similar algebra in \eqref{deltadAU}, from \eqref{pvfRdA} and \eqref{weight} we find
\begin{align}
\label{entropy expansion}
f'(R)\td A (v, \mathcal{U}, x^i) &= f'(R) \td A (v, 0, x^i) + \mathcal{U}(v, x^i) \partial_u\left(f'(R) \td A (v, 0, x^i)\right)\\
&= (1-v\partial_{v})f'(R)\td A\,(v,0,x^{i})-v\td A\,(v,0,x^{i})D_{i}(f'(R)(D^{i}{\cal U}+\omega^{i}{\cal U}))\nn\,. 
\end{align}
When integrating over the codimension-2 surface, the second term on the right-hand side as a total derivative will not contribute. Then, we obtain
\begin{equation}
    \int f'(R)\td A(v,\mathcal{U})=(1-v\partial_{v})\int f'(R)\td A(v,0)\,.
\end{equation}
The left-hand side of the above equation can be recognized as the integral of $f'(R)$ on $\tilde{\cal T}$, and hence \eqref{SdynSwaldfR} can be written as
\begin{equation}
\label{SdynfR}
    S_{\text{dyn}}=S_{\text{Wald}}[\tilde{\mathcal{T}}(v)]=\frac{1}{4G}\int_{\tilde{\cal T}}f'(R)\,.
\end{equation}
Therefore, to the first order in perturbations around a stationary black hole background, the dynamical black hole entropy in $f(R)$ gravity equals the Wald entropy evaluated on the generalized apparent horizon.

\section{Entanglement Entropy for Dynamical Black Holes}
\label{Gravitational Entanglement Entropy Around Different Surfaces} 

When interpreting the thermal entropy of a stationary black hole, the bifurcation surface serves as a natural entangling surface because it lies at the boundary between causally disconnected regions of spacetime. However, the situation becomes subtle for dynamical black holes, where the event horizon and apparent horizon no longer coincide. From \eqref{2.45} and \eqref{SdynfR}, we have seen that for Einstein's gravity and $f(R)$ gravity, the dynamical black hole entropy under the first order in the perturbation becomes the Bekenstein-Hawking/Wald entropy of the apparent horizon. This strongly suggests that for dynamical black holes, the apparent horizon $\mathcal{T}(v)$ represents the true physical horizon rather than the future event horizon $\mathcal{H}^+$. The physical interpretation of this becomes clear when considering the null expansion $\tilde{\theta}_{\tilde{k}}$ at the apparent horizon $\mathcal{T}(v)$. The condition $\tilde{\theta}_{\tilde{k}}=0$ at $\mathcal{T}(v)$ indicates that it is a marginal surface where outgoing light rays neither converge nor diverge, which can be naturally viewed as a defining characteristic of a horizon for dynamical situations.

In this section, we will explicitly compute the entropy for a dynamical black hole using the replica method, taking both the event and apparent horizons as candidate entangling surfaces. We will show that only the apparent horizon yields an entropy that matches the thermal entropy $ S_{\text{dyn}} $.  

\subsection{Gravitational Entropy from the Replica Method}
\label{entanglement entropy}
In this subsection, we briefly review the calculation of entanglement entropy using the replica trick in the path integral formalism. For a comprehensive review, the reader may refer to, e.g., \cite{Nishioka:2018khk}. 

We start with the entanglement entropy in quantum field theory. Suppose a Cauchy surface is divided into two complementary spatial regions $\rm A$ and $\rm B$. The quantum state $|\Psi\rangle$ is prepared via the Euclidean path integral from the Euclidean time $\tau = -\infty$ to $\tau = 0$, representing the vacuum state. Field configurations on $\Sigma$ are denoted by $\phi(\Sigma)$, with $\phi^{\rm A}$ and $\phi^{\rm B}$ indicating field values restricted to regions $\rm A$ and $\rm B$, respectively. The reduced density matrix for region $\rm A$ is obtained by tracing out degrees of freedom in $\rm B$:
\begin{equation}
\rho_{\text{A}}=\frac{1}{Z}\int[\mathcal{D}\phi^{\text{B}}(\Sigma)]\langle \phi^{\text{B}} |\Psi \rangle \langle \Psi |\phi^{\text{B}} \rangle\,,
\end{equation}
where $|\phi^B\rangle$ are eigenstates of the field operator $\hat{\phi}(x)$ in region $\rm B$. The normalization factor $Z$ is given by the complete path integral:
\begin{equation}
Z=\int[\mathcal{D}\phi(\Sigma)]\langle \Psi |\phi \rangle \langle \phi |\Psi \rangle\,.
\end{equation}
The wavefunctional $\langle \phi |\Psi \rangle $ computes the overlap between the field configuration $\phi(\Sigma)$ and the vacuum state:
\begin{equation}
\langle \phi |\Psi \rangle =\int^{\Sigma,\phi}_{\tau=-\infty}[\mathcal{D}\phi]e^{-I_{E}[\phi]}\,.
\end{equation}
The entanglement entropy can be evaluated by first introducing the R\'enyi entropy:
\begin{equation}
S_n(\text{A}) = \frac{1}{1-n} \log \text{tr} (\rho_{\rm A}^n)\,.
\end{equation}
Then, entanglement entropy $S_{\rm A}$ is obtained through analytic continuation in the replica index $n$:
\begin{equation}
S_{\text{A}}=\lim_{n\rightarrow 1}S_{n}(\text{A})=-\lim_{n\rightarrow 1}\partial_{n}\log \text{tr}_{\text{A}}(\rho^{n}_{\text{A}})\,.
\end{equation}
This approach, known as the replica trick, provides a powerful method for computing entanglement entropy in quantum field theories.

In gravitational systems, the Euclidean path integral includes the integration over both matter fields $\phi$ and the metric $g_{\mu\nu}$, with the general form:
\begin{equation}
Z = \int [\mathcal{D}g][\mathcal{D}\phi] \, e^{-I_E[g,\phi]}\,,
\end{equation}
where $I_E[g,\phi]$ is the Euclidean Einstein-Hilbert action coupled to matter fields. However, this path integral is notoriously ill-defined due to the non-renormalizable nature of quantum gravity and the unboundedness of the Euclidean gravitational action. In practice, what one can do is to employ the saddle-point approximation around a classical solution $\bar g_{\mu\nu}$, e.g., a Schwarzschild black hole. The dominant contribution comes from the on-shell action reads (later on, we will drop the bar of $\bar g$ for simplicity)
\begin{equation}
Z \simeq e^{-I_E[\bar g]}\,.
\end{equation}
For entropy calculations, the replica trick requires constructing an $n$-fold branched cover $\mathcal{M}_n$ of the original spacetime $\mathcal{M}$, with branch cuts along the entangling surface $\cal S$ (e.g., the horizon of a black hole). The gravitational entropy across $\cal S$ is then extracted from the partition function's dependence on $n$:
\begin{align}
\label{SEE}
S_{\text{grav}} &= \lim_{n\to1}\frac{1}{1-n} \left( \log Z_n - n \log Z \right)\\
\label{entanglement entropy calculation formula}
&= -\lim_{n\to1}\partial_n \left( \log Z_n - n \log Z \right)\,,
\end{align}
where $Z_n$ is the partition function evaluated semiclassically on ${\cal M}_n$. For a stationary black hole, it is well-known that this result matches the Bekenstein-Hawking entropy $S_{\text{BH}} = \frac{A}{4G}$ when applied to the black hole horizon.

Before turning to dynamical black holes, we emphasize that the replica trick computes the entropy of a reduced density matrix defined with respect to a choice of state $|\Psi\rangle$ and a choice of entangling surface on a Cauchy slice. For generic dynamical black holes there is no canonical analogue of the Hartle-Hawking state, and a global Euclidean section is generally complex and not unique. In this work, we do not attempt to define a nonperturbative Euclidean geometry or a preferred state for an arbitrary fully dynamical spacetime. Instead, we work perturbatively around a stationary black hole background for which $|\Psi\rangle$ is prepared by the Hartle-Hawking Euclidean path integral, and we extract the entropy functional from \eqref{entanglement entropy calculation formula} using a regulated conical defect localized near the candidate surface. Our goal in the rest of this section is to compare different candidate surfaces and identify which choice reproduces the dynamical black hole entropy $S_{\rm dyn}$ that satisfies the physical process first law.

\subsection{Across the Future Event Horizon}
\label{Sec:Future Event Horizon}

We begin by computing the gravitational entropy functional \eqref{entanglement entropy calculation formula} with the future event horizon taken as a candidate entangling surface.

Working in the GNC gauge, we keep fixed the affine parameters $v$ (for $\mathcal{H}^+$) and $u$ (for $\mathcal{H}^-$), as well as the spatial coordinates $x^i$, under perturbations. This means any non-stationary perturbation only affects the metric functions, as shown in \eqref{dynamical perturbation}. To analyze the entropy associated with a cross section ${\cal C}$ of the future horizon at affine time $v_{0}$, we perform the following coordinate transformation:
\begin{align}
    u&=\frac{1}{\sqrt{2}}(r\sinh\tau-r\cosh\tau)\,,\\
    v&=v_{0}+\frac{1}{\sqrt{2}}(r\sinh\tau+r\cosh\tau)\,.
\label{coordinate transformation 1}
\end{align}
In the coordinate system $(\tau,r,x^i)$, the line element \eqref{2.7} now reads
\begin{equation}
\label{metricFEH}
    \td s^{2}=(1-\frac{\alpha}{4}r^{2})\td r^{2}-r^{2}(1+\frac{\alpha}{4}r^{2})\td\tau^{2}-\frac{\alpha}{2}r^{3}\td r\td\tau+r\omega_{i}\td r\td x^{i}+r^{2}\omega_{i}\td\tau \td x^{i}+\gamma_{ij}\td x^{i}\td x^{j}\,,
\end{equation}
where functions $\alpha(\tau,r,x^i)$, $\omega_{i}(\tau,r,x^i)$ and $\gamma_{ij}(\tau,r,x^i)$ are already perturbed. Note that in the new coordinate system, these metric functions are not singular at $r=0$. In this system, the entangling surface is located at $r=0$.

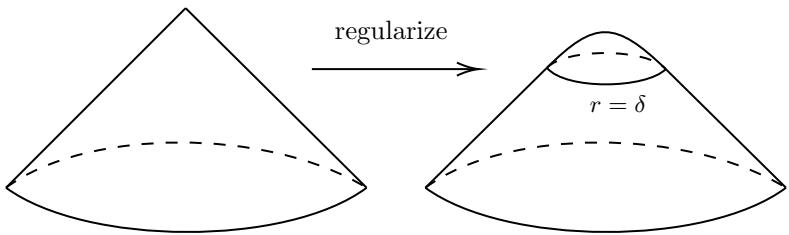
\begin{figure}[!h]
    \centering

\tikzset{every picture/.style={line width=0.75pt}} 

\begin{tikzpicture}[x=0.75pt,y=0.75pt,yscale=-1,xscale=1]

\draw    (160.33,79.67) -- (70.67,170) ;
\draw    (160.33,79.67) -- (250.67,170) ;
\draw    (70.67,170) .. controls (110.67,199.33) and (210.67,200) .. (250.67,170) ;
\draw  [dash pattern={on 4.5pt off 4.5pt}]  (70.67,170) .. controls (110.67,140) and (200.67,139.33) .. (250.67,170) ;
\draw    (340.67,109.67) -- (280,170) ;
\draw    (400,110.33) -- (460,170) ;
\draw    (280,170) .. controls (320,199.33) and (420,200) .. (460,170) ;
\draw  [dash pattern={on 4.5pt off 4.5pt}]  (280,170) .. controls (320,140) and (410,139.33) .. (460,170) ;
\draw  [dash pattern={on 4.5pt off 4.5pt}]  (340.67,109.67) .. controls (350,99.67) and (390.67,99.67) .. (400,110.33) ;
\draw    (340.67,109.67) .. controls (350,121) and (390,120.33) .. (400,110.33) ;
\draw    (340.67,109.67) .. controls (364.67,85.67) and (374.67,85.67) .. (400,110.33) ;
\draw    (223.33,110.33) -- (304.67,110.33) ;
\draw [shift={(306.67,110.33)}, rotate = 180] [color={rgb, 255:red, 0; green, 0; blue, 0 }  ][line width=0.75]    (10.93,-3.29) .. controls (6.95,-1.4) and (3.31,-0.3) .. (0,0) .. controls (3.31,0.3) and (6.95,1.4) .. (10.93,3.29)   ;

\draw (376.25,127.53) node  [font=\footnotesize]  {$r=\delta $};
\draw (233.33,84.67) node [anchor=north west][inner sep=0.75pt]  [font=\small] [align=left] {regularize};

\end{tikzpicture}
\caption{The conical singularity at $r= 0$ is regularized by introducing the smooth function $\beta_\delta$.}
\label{fig3}
\end{figure}

We evaluate the gravitational entropy \eqref{entanglement entropy calculation formula} by Wick rotating $\tau\to i t$ and constructing the $n$-fold cover $\mathcal M_n$., on which $r\geqslant0$ and $0\leqslant t\leqslant 2\pi n$. In order to properly take into account the conical singularity at the origin, we regularize the cone using the smooth function $\beta_\delta(r)$ satisfying 
\begin{align}
\label{regulatorbeta}
    \beta_\delta(r \to 0) = n^2\,,\qquad \beta_\delta(r > \delta) = 1\,,
\end{align}
where $\delta$ is an arbitrary regularization parameter independent of any UV cutoff, which makes $\tilde{C}_n$ smooth at $r=0$ while matching $C_n$ for $r > \delta$, as shown in Figure~\ref{fig3}. We emphasize that the replica construction is used here in a local, near-surface manner: the regulator $\beta_\delta(r)$ isolates the contribution supported in the transition region $r\sim\delta$ near the tip, which controls the $n\to1$ limit in \eqref{entanglement entropy calculation formula}. This calculation does not require specifying a unique global Euclidean completion of the full dynamical geometry; its main purpose is to compare candidate surfaces and test which choice reproduces the entropy $S_{\text{dyn}}$ satisfying the physical process first law.

After regularization, the line element of the metric on $\mathcal{M}_{n}$ becomes
\begin{equation}
    \td s_{\mathcal{M}_{n}}^{2}=(1-\frac{\alpha}{4}r^{2})\beta_{\delta}\td r^{2}+r^{2}(1+\frac{\alpha}{4}r^{2})\td t^{2}-\I\frac{\alpha}{2}r^{3}\td r\td t+r\omega_{i}\td r\td x^{i}+\I r^{2}\omega_{i}\td t \td x^{i}+\gamma_{ij}\td x^{i}\td x^{j}\,.
    \label{line element of event horizon}
\end{equation}
Although the Wick-rotated metric is generally complex away from the tip for a non-stationary perturbation, the derivative at $n=1$ appearing in \eqref{entanglement entropy calculation formula} is determined by the local conical structure in the $(r,t)$ directions near $r=0$, encoded in $g_{rr}\sim \beta_\delta(r)$ and $g_{tt}\sim r^2$. While the Wick-rotated metric also contains complex cross terms, what controls the entropy functional is the localized contribution in the curvature near the tip. As shown explicitly in Appendix~\ref{Details of entanglement entropy calculation}, the cross terms do not contribute to this localized piece at the order we work. Accordingly, we treat the replica computation as a saddle-point evaluation of the Euclidean action on a geometry with a regulated conical defect, and we do not assume the existence of a globally real Euclidean continuation of the full perturbed spacetime.\footnote{When varying the Euclidean action $I_E$, the parameters $\alpha$, $\omega_i$, and $\gamma_{ij}$ appearing in the metric are treated as real. Hence, the saddle condition $\delta I_E = 0$ implies both $\delta\,\mathrm{Re}[I_E] = 0$ and $\delta\,\mathrm{Im}[I_E] = 0$ for the allowed variations. Since the semiclassical weight is controlled by $\mathrm{Re}[I_E]$, while $\mathrm{Im}[I_E]$ contributes a phase, the saddle-point approximation remains well defined in our perturbative treatment.}

To calculate the partition function $Z_{n}$ on $\mathcal{M}_{n}$ in the semiclassical limit, we need to consider the on-shell gravitational action. To begin with, we consider the Euclidean Einstein-Hilbert action
\begin{equation}
\label{EHaction}
    I_E=-\frac{1}{16\pi G}\int_{\mathcal{M}_{n}}\td^{d}x\sqrt{\det g}R\,.
\end{equation}
Since the metric is symmetric, we can use the Schur complement formula to calculate the determinant. For a $2\times 2$ block matrix
\begin{equation}
    g=\begin{pmatrix}D&B\\ B^{T}&\gamma\end{pmatrix},
\end{equation}
the determinant is
\begin{equation}
    \det g=\det D\cdot\det(\gamma-B^{T}D^{-1}B)
    \label{Schur complement formula}
\end{equation}
if $D$ is an invertible matrix. In our case $D$ is obviously invertible, then we can obtain from \eqref{line element of event horizon} that
\begin{equation}
    \det D=r^2\Big(\beta_{\delta}+\frac{\alpha^{2}}{16}(1-\beta_{\delta})r^{4}\Big)\,.
\end{equation}
When $r>\delta$, we have $\beta_{\delta}=1$ as defined in \eqref{regulatorbeta}. On the other hand, since $\delta$ is small, the second term is at least ${\cal O}(r^4)$, and thus is negligible when $r<\delta$. Therefore, for the whole spacetime, we have
\begin{equation}
    \det D=r^2\beta_{\delta}\,.
\end{equation}
Furthermore, from \eqref{line element of event horizon} we also find that
\begin{equation}
    (B^{T}D^{-1}B)_{ij}=\frac{1}{4}\bigg(\frac{1+\alpha r^{2}/4}{\beta}- (1-\alpha r^{2}/4)-\frac{\alpha r^{4}}{2}\bigg)r^{2}\omega_{i}\omega_{j}\,.
    \label{BAB}
\end{equation}
This term is at least ${\cal O}(r^2)$ near $r=0$. Therefore, when we approach the entangling surface at $r=0$, we have
\begin{equation}
\label{detgapprox}
    \det g=r^{2}{\beta_{\delta}}\det(\gamma-B^{T}D^{-1}B)\simeq r^{2}{\beta_{\delta}}\det\gamma\,.
\end{equation}

Finally, we now calculate the Ricci scalar in \eqref{EHaction}. From \eqref{line element of event horizon}, one can find that the Ricci scalar can be organized into a part without the $r$-derivative of $\beta_{\delta}$, denoted by $R_0$, and a part proportional to the $r$-derivative of $\beta_{\delta}$. The integral of $R_0$ over ${\cal M}_n$ is proportional $n$ times the integral of the Ricci scalar on the original geometry $({\cal M},g)$ (see Appendix~\ref{Details of entanglement entropy calculation}). Therefore,
\begin{equation}
\label{intR}
\int _{\mathcal{M}_{n}}R=\int _{\mathcal{M}_n}R_0+\int_{\mathcal{M}_{n}}\eta\beta'_{\delta}=n\int _{\mathcal{M}}R+\int_{\mathcal{M}_{n}}\eta\beta'_{\delta}\,,
\end{equation}
where $\beta'_{\delta}$ represents the $r$-derivative of $\beta_\delta$. The second term on the right-hand side only receives the contribution at $r<\delta$. As analyzed in Appendix~\ref{Details of entanglement entropy calculation}, near the future horizon we can see that $\eta\simeq \frac{1}{r\beta_{\delta}^2}$, and hence the second integral reads
\begin{align}
\int_{\mathcal{M}_{n},r<\delta}\td r\td t\td^{d-2}xr\sqrt{\beta_{\delta}}\sqrt{\det\gamma}\frac{\beta'_{\delta}}{r\beta_{\delta}^{2}}=-2\int_{\mathcal{M}_{n},r<\delta}\td r\td t\td^{d-2}x\sqrt{\det\gamma}\partial_{r}(\beta_{\delta}^{-\frac{1}{2}})\,,
\label{A.14}
\end{align}
where we have used the determinant in \eqref{detgapprox}. In the limit $\delta\to0^+$, we can show that
\begin{equation}
    \beta_{\delta}^{-\frac{1}{2}}=\frac{1}{n}+\Big(1-\frac{1}{n}\Big)\theta(r-0^{+})\,.
    \label{A.15}
\end{equation}
Then, from \eqref{intR} we have
\begin{equation}
\int _{\mathcal{M}_{n}}R-n\int _{\mathcal{M}}R=\frac{2(1-n)}{n}\int_{\mathcal{M}_{n},r<\delta}\td r\td t\td^{d-2}x\sqrt{\det\gamma}\delta(r)=4\pi(1-n)A[\mathcal{C}]\,.
\label{results in future horizon}
\end{equation}
Plugging the result \eqref{results in future horizon} into the formula \eqref{entanglement entropy calculation formula}, we obtain
\begin{equation}
    S_{\text{grav}}[\mathcal{C}(v)]=\frac{1}{4G}A[\mathcal{C}(v)]\,.
\end{equation}
Therefore, for a black hole with non-stationary perturbation, the entropy across the future event horizon still satisfies the area law of a section ${\cal C}(v)$. However, as we have seen in Section~\ref{Dynamical Black Hole Entropy from Raychaudhuri Equation}, the thermal entropy $S_{\rm dyn}$ for a dynamical black receives a correction from the perturbation, and thus does not match the entropy $S_{\text{ent}}[\mathcal{C}(v)]$. It is also obvious that $S_{\text{ent}}[\mathcal{C}(v)]$ does not satisfy the physical process first law.

The above result for Einstein's gravity can also be generalized to $f(R)$ gravity. Consider the following effective action on ${\cal M}_n$:
\begin{equation}
\label{fRaction}
    I_E=-\frac{1}{16\pi G}\int_{\mathcal{M}_{n}}\td^{d}x\sqrt{\det g}f(R)\,.
\end{equation}
When evaluating the integrand $f(R)$, we can still express it as $f(R_0+\eta\beta_\delta')$. Near the horizon, we have in the limit $\delta\to 0^+$,
\begin{align}
\lim_{\delta\to 0^+}\eta\beta_\delta'=\lim_{\delta\to 0^+}-\frac{2}{r\sqrt{\beta_\delta}}\partial_r(\frac{1}{\sqrt{\beta_\delta}})=\frac{2(1-n)}{n}\frac{\delta(r)}{r\sqrt{\beta_\delta}}\,.
\end{align}
On the other hand, we notice that the integral of $f(R_0)$ is equal to $n$ times the integral of $f(R)$ on the original geometry $({\cal M},g)$ when $\delta\to 0^+$, i.e.,
\begin{align}
\int_{\mathcal{M}_n}f(R_0)=n\int_{\mathcal{M}}f(R)\,.
\end{align}
Then, for $f(R)$ gravity, the replica entropy equation \eqref{SEE} can be written as
\begin{align}
S_{\text{grav}}&=\frac{1}{16\pi G}\lim_{n\to1}\frac{\int_{\mathcal{M}_n}f(R_0+(1-n)\delta R)-\int_{\mathcal{M}_n}f(R_0)}{1-n}\nn\\
\label{SEEfR}
&=\frac{1}{16\pi G}\lim_{n\to1}\frac{\int_{\mathcal{M}_n,r<\delta}f(R_0+(1-n)\delta R)-\int_{\mathcal{M}_n,r<\delta}f(R_0)}{1-n}\,,
\end{align}
where $\delta R\equiv\frac{2}{n}\frac{\delta(r)}{r\sqrt{\beta_\delta}}$. Denote $F[R]=\int_{\mathcal{M}_n,r<\delta}f(R)$, then the limit in the above expression gives the functional derivative of $F[R]$. It is straightforward to compute that
\begin{align}
\delta F[R]=\int_{\mathcal{M}_n,r<\delta}f'(R)\delta R=4\pi\int_{\mathcal{C}}f'(R)\,.
\end{align}
Therefore, the entropy across the future event horizon is
\begin{equation}
     S_{\text{grav}}[\mathcal{C}(v)]=S_{\text{Wald}}[{\cal C}(v)]=\frac{1}{4G}\int_{\mathcal{C}}f'(R)\,.
\end{equation}
This indicates that when taking ${\cal C}$ as the entangling surface, the entropy $S_{\rm grav}$ for $f(R)$ gravity still agrees with the standard Wald entropy of the unperturbed geometry but does not match $S_{\rm dyn}$.

\subsection{Across the Apparent Horizon}
\label{Sec:Apparent Horizon}

To calculate the gravitational entropy across the apparent horizon, we define a pair of new parameters $\tilde{u}=u-\mathcal{U}(v_{0},x^i)$ and $\tilde{v}=v$. We extend $\tilde{k}^{a}$ off the apparent horizon as follows:
\begin{equation}
        \tilde{k}^{a}=(\partial_{\tilde{v}})^{a}+(D^{i}\mathcal{U}+\omega^{i}\mathcal{U})(\partial_{i})^{a}\,.
    \label{4.26}
\end{equation}
In the coordinate system $(\tilde u,\tilde v,x^i)$, the position of the apparent horizon is at $\tilde{u}=0$.
We will focus on the geometry near a cross section $\mathcal{T}$ of the apparent horizon. In the new coordinates, the line element of the metric can be written as
\begin{equation}
\td s^{2}=-2\td\tilde{u}\td\tilde{v}+g_{\tilde{v}\tilde{v}}\td\tilde{v}^{2}+g_{\tilde{v}i}\td\tilde{v}\td x^{i}+\gamma_{ij}\td x^{i}\td x^{j}+\mathcal{O}(\epsilon^{2})\,,
\label{4.27}
\end{equation}
where $g_{\tilde{v}\tilde{v}}=\tilde{u}b(\tilde{v},\tilde{u},x^{i})=\tilde{u}(2\mathcal{U}+\tilde{u}\alpha)$ and $g_{\tilde{v}i}=-2D_{i}\mathcal{U}-2\mathcal{U}\omega_{i}-2\tilde{u}\omega_{i}+\mathcal{O}(\epsilon^{2})$.  
We can find that the corresponding dual vector for $\tilde{k}^{a}$ reads
\begin{equation}
    \tilde{k}_{a}=-(\td \tilde{u})_{a}-\tilde{u}b(\td \tilde{v})_{a}-\tilde{u}\omega_{i}(\td x^{i})_{a}\,.
\end{equation}
At the apparent horizon, we have $\tilde{k}_{a}=-(\td \tilde{u})_{a}=-(\td u)_{a}+\partial_{i}\mathcal{U}(\td x^{i})_{a}$, which agrees with \eqref{2.26}. It is also easy to show that $\tilde{l}^{a}=(\partial_{\tilde{u}})^{a}=(\partial_{u})^{a}=l^{a}$. 

Applying the following coordinate transformation:
\begin{align}
\tilde{u}&=\frac{1}{\sqrt{2}} \tilde{r} \left( \sinh \tilde{\tau} -\cosh \tilde{\tau} \right)\,,\\
    \tilde{v} &=\tilde{v}_{0} +\frac{1}{\sqrt{2}} \tilde{r} \left( \sinh \tilde{\tau} +\cosh \tilde{\tau} \right)\,,
\end{align}
as well as the Wick rotation $\tilde{\tau}\rightarrow \I\tilde{t}$, we get the Euclidean metric. Then, the line element on the $n$-fold cover $\mathcal{M}_{n}$ of the spacetime with $\tilde{r}\geqslant0$ and $0\leqslant \tilde{t}\leqslant 2\pi n$ becomes (up to first order of $\epsilon$)
\begin{equation}
\td s^{2}=g_{\tilde{r}\tilde{r}}\beta_{\delta}(\tilde{r})\td\tilde{r}^{2}+g_{\tilde{r}\tilde{t}}\td\tilde{r}\td\tilde{t}+g_{\tilde{t}\tilde{t}}\td\tilde{t}^{2}+g_{\tilde{r}i}\td\tilde{r}\td x^{i}+g_{\tilde{t}i}\td\tilde{t}\td x^{i}+\gamma_{ij}\td x^{i}\td x^{j}\,,
\label{line element of the apparent horizon}
\end{equation}
where $g_{\tilde{r}\tilde{r}}=1-b'\tilde{r}$, $g_{\tilde{r}\tilde{t}}=-2\I b'\tilde{r}^{2}$, $g_{\tilde{t}\tilde{t}}=\tilde{r}^{2}(1+b'\tilde{r})$, $g_{\tilde{r}i}=\mathcal{O}(\epsilon)+\tilde{r}\omega_{i}$ and $g_{\tilde{t}i}=\I\tilde{r}g_{\tilde{r}i}$, with $b'=\frac{1}{2\sqrt{2}} b\E^{\I \tilde t}$.
 Again, the smooth function $\beta_{\delta}(\tilde r)$ is introduced for regularizing the conical singularity at the origin, satisfying
\begin{equation}
    \beta_{\delta}(\tilde{r}\rightarrow0)=n^{2}\,,\qquad \beta_{\delta}(\tilde{r}>\delta)=1\,.
\end{equation} 
Following a similar analysis as the previous subsection, we can see that
\begin{equation}
    \det g=\tilde r^2{\beta_{\delta}}\det(\gamma-{\cal O}(\epsilon^2))\simeq \tilde r^2{\beta_{\delta}}\det\gamma\,.
\end{equation}

For Einstein's gravity, we should evaluate the integral of the Ricci scalar in a similar manner as \eqref{intR}, except now the entangling surface is the apparent horizon at $\tilde r=0$, and the relevant term is the proportional to $\beta_\delta'(\tilde r)$. Going through the same calculation as the event horizon case, we arrive at (see Appendix~\ref{Details of entanglement entropy calculation})
\begin{equation}
\label{results in apparent horizon}
    \int_{\mathcal{M}_{n}}R-n\int_{\mathcal{M}} R=4\pi(1-n)A[\mathcal{T}]+{\cal O}(\epsilon^2)\,.
\end{equation}
Plugging \eqref{results in apparent horizon} into formula \eqref{entanglement entropy calculation formula}, up to first order in the perturbation we get
\begin{equation}
    S_{\text{grav}}[\mathcal{T}(v)]=\frac{1}{4G}A[\mathcal{T}(v)]\,.
\end{equation}
We can see that this result matches exactly the entropy $S_{\text{dyn}}$ in \eqref{2.45} under a non-stationary perturbation.

We now extend our analysis to $f(R)$ gravity by evaluating the entropy across the generalized apparent horizon $\tilde{\mathcal{T}}$. As established in Section~\ref{Dynamical Black Hole Entropy from Raychaudhuri Equation}, the near-horizon geometry of $\tilde{\mathcal{T}}(v)$ is the same as that of the apparent horizon in \eqref{metric decomposition}, with the only difference being the horizon location function $\mathcal{U}$. This means the geometric framework set up in \eqref{4.26}, \eqref{4.27}, and \eqref{line element of the apparent horizon} continues to hold when using the $\mathcal{U}$ associated with the generalized apparent horizon. Following the same evaluation for $f(R)$ gravity as outlined in \eqref{SEEfR}, up to first order in the perturbation we arrive at
\begin{equation}
S_{\text{grav}}[\tilde{\mathcal{T}}(v)] = S_{\text{Wald}}[\tilde{\mathcal{T}}(v)] = \frac{1}{4G}\int_{\tilde{\mathcal{T}}} f'(R)\,,
\end{equation}
which precisely agrees with the entropy $S_{\text{dyn}}$ in \eqref{SdynfR} derived for $f(R)$ gravity. 

The results in this section demonstrate that the entropy $S_{\rm grav}$ computed across both the future event horizon and the apparent horizon reproduces the Bekenstein-Hawking entropy (in Einstein gravity) and the Wald entropy (in $f(R)$ gravity) associated with the corresponding surfaces. However, under non-stationary perturbations, only the $S_{\rm grav}$ across the apparent horizon matches the corrected thermal entropy $S_{\text{dyn}}$ and satisfies the physical process first law. These distinctions arise because the future event horizon, being teleological in nature, does not reflect local thermodynamic equilibrium in dynamical scenarios, whereas the apparent horizon, defined by $ \tilde{\theta}_{\tilde{k}} = 0 $, represents the instantaneous physical boundary of the black hole.

\section{Modified von Neumann Entropy and the Generalized Second Law}
\label{Modified von Neumann Entropy and the Generalized Second Law}

Using the quantum null energy condition (QNEC), the validity of the generalized second law (GSL) for a black hole with non-stationary perturbation was verified in \cite{Hollands:2024vbe} by means of the notion of modified von Neumann entropy. In this section, we further investigate the GSL and provide a physical interpretation for the modified von Neumann entropy. 

For a quantum field with stress-energy tensor $T_{ab}$ on a fixed classical background, the QNEC, which was first introduced in \cite{Bousso:2015mna}, says that on a cross section $\mathcal{C}$ of a null hypersurface $\mathcal{N}$, where the expansion and shear of $\mathcal{N}$ vanish, the von Neumann entropy $S_{\text{vN}}$ of the quantum field outside\footnote{Here, ``outside'' means the region on a spacelike hypersurface crossing $\cal C$, whose future Cauchy horizon coincides with $\mathcal{N}$.} the null hypersurface satisfies
\begin{equation}
\label{QNEC}
    \langle T_{ab}\rangle k^{a}k^{b}\geqslant \frac{1}{2\pi A[{\cal C}]}\frac{\delta^{2}}{\delta v(x)^{2}}S_{\text{vN}}\,.
\end{equation}
In the black hole geometry we considered, ${\cal H}^+$ plays the role of $\cal N$. In general, the variation can be performed differently at each point $x$ on $\cal C$. For simplicity, we will only consider uniform deformations in $v$, which does not depend on $x$.  Then, integrating both sides of \eqref{QNEC}, we obtain
\begin{equation}
    \int_{\mathcal{C}}\langle T_{ab}\rangle k^{a}k^{b}\geqslant \frac{1}{2\pi}\frac{\td^{2}}{\td v^{2}}S_{\text{vN}}\,.
    \label{4.18}
\end{equation}

In Section~\ref{Dynamical Black Hole Entropy from Raychaudhuri Equation}, we have seen from \eqref{perturbedRaychaudhuri} and \eqref{Sdynderive} that the Raychaudhuri equation and the Einstein equation give
\begin{equation}
  \frac{\kappa}{2\pi}\frac{\td}{\td v}\delta S_{\text{dyn}}=\int_{\mathcal{C}}\delta\langle T_{ab}\rangle k^{a}\xi^{b}=\kappa v\int_{\mathcal{C}}\delta\langle T_{ab}\rangle k^{a}k^{b}\,.
\end{equation}
Applying the QNEC \eqref{4.18}, we obtain
\begin{equation}
    \frac{\td}{\td v}\delta S_{\text{dyn}}\geqslant v\frac{\td^2}{\td v^2}\delta S_{\text{vN}}\,.
    \label{4.20}
\end{equation}
In fact, it can be verified that the above equation holds for any diffeomorphism invariant theory \cite{Hollands:2024vbe, Visser:2024pwz}. To obtain the GSL, one can define the \emph{modified von Neumann entropy} as follows:
\begin{equation}
\tilde{S}_{\text{vN}}=S_{\text{vN}}-v\frac{\td}{\td v}S_{\text{vN}}\,.
\end{equation}
It is obvious that $\tilde{S}_{\text{vN}}$ agrees with $S_{\text{vN}}$ in the stationary case, and the second term provides a correction when $S_{\text{vN}}$ changes with the affine time $v$. The derivative of the modified von Neumann entropy reads
\begin{equation}
\label{dSvN}
\frac{\td}{\td v}\tilde{S}_{\text{vN}}=-v\frac{\td^2}{\td v^2} S_{\text{vN}}\,.
\end{equation}
Combining \eqref{4.20} and \eqref{dSvN} yields
\begin{equation}
    \frac{\td}{\td v}(\delta S_{\text{dyn}}+\delta \tilde{S}_{\text{vN}})\geqslant 0\,,
    \label{The Generalized Second Law}
\end{equation}
which is recognized as the GSL of a dynamical black hole \cite{Hollands:2024vbe}.

Following our observation that the apparent horizon serves as a natural entangling surface in the dynamical case, we now discuss the physical interpretation of the modified von Neumann entropy. For simplicity, we first restrict our consideration to Einstein’s gravity. By definition, the von Neumann entropy of the matter field outside the black hole is  
\begin{equation}  
S_{\text{vN}} = -\text{tr}(\rho_{\text{out}} \log \rho_{\text{out}})\,,  
\end{equation}  
where $\rho_{\text{out}}$ is the reduced density matrix of the quantum matter obtained by taking $\mathcal{C}(v)$ as the entangling surface. For a quantum field theory on curved spacetime, $S_{\text{vN}}$ can be evaluated using the effective action via the replica trick. Integrating out the matter field in the path integral yields a diffeomorphism-invariant effective action composed of curvature polynomials:  
\begin{equation}
\label{IEmatter}
I_{\text{E, matter}} = \int \mathrm{d}^d x \sqrt{\det g} ( c_2 \Lambda^{d-2} R + c_{4,1} \Lambda^{d-4} R^2 + c_{4,2} \Lambda^{d-4} R_{\mu\nu}^2 + c_{4,3} \Lambda^{d-4} R_{\mu\nu\rho\sigma}^2 + \cdots)\,,  
\end{equation}  
where $\Lambda$ is the UV cutoff scale. Here, we omit the $\Lambda^d$ term as it only contributes to the cosmological constant. In even dimensions, there is also an additional logarithmic divergence, which gives rise to the Weyl anomaly. For now, let us consider matter fields in the Hawking-Hartle vacuum and focus on the leading divergence, namely the Einstein-Hilbert term $c_2 \Lambda^{d-2} R$ in the effective action.

The leading UV-divergent local contribution to the matter entanglement entropy can be extracted from the conical variation of the matter effective action. As we have seen in Section~\ref{Gravitational Entanglement Entropy Around Different Surfaces}, the conical geometry implies that by choosing a cross section $\cal C$ of ${\cal H}^+$ as the entangling surface and applying the replica trick, we get
\begin{equation}
    \int_{\mathcal{M}_{n}}R-n\int_{\cal M} R=4\pi(1-n)A[{\cal C}]\,.
\end{equation}
For minimally coupled matter, the leading divergence is controlled by the induced Einstein-Hilbert term $c_2 \Lambda^{d-2} R$, whose conical variation gives an area-law contribution proportional to $A[\mathcal{C}(v)]$:
\begin{equation}
    S_{\text{vN}}=-4\pi c_{2}\Lambda^{d-2}A[\mathcal{C}(v)]\,.
\end{equation}
We stress that for gauge fields and for non-minimally coupled scalars there are additional subtleties---such as contact term/edge-mode contributions \cite{Donnelly:2012st}---and the simple identification of the entropy with the conical variation of local bulk terms requires refinement; our discussion here focuses only on the leading local area divergence. Then, it follows from \eqref{2.17} and \eqref{2.45} that
\begin{equation}
    S_{\text{vN}}-v\frac{\td}{\td v}S_{\text{vN}}=-4\pi c_{2}\Lambda^{d-2}(1-v\theta_{v})A[\mathcal{C}(v)]=-4\pi c_{2}\Lambda^{d-2}A[\mathcal{T}(v)]\,.
\end{equation}
Therefore, at the level of the leading local area-law divergence, we may interpret the modified von Neumann entropy as the matter entanglement entropy evaluated across the apparent horizon, i.e.,
\begin{equation}
    \tilde{S}_{\text{vN}}=S_{\text{matter}}[\mathcal{T}(v)]\,.
\end{equation}

In Section~\ref{Gravitational Entanglement Entropy Around Different Surfaces} we identified the dynamical black hole entropy $S_{\rm dyn}$ as the gravitational entropy $S_{\rm grav}$ across the apparent horizon, then \eqref{The Generalized Second Law} can be written as
\begin{equation}
    \frac{\td}{\td v}(\delta S_{\text{grav}}[\mathcal{T}(v)]+\delta S_{\text{matter}}[\mathcal{T}(v)])\geqslant 0\,.
    \label{4.34}
\end{equation}
This explicitly justifies the physical meaning of \eqref{The Generalized Second Law} as the generalized second law: the total entropy of system, including the gravitational contribution and the matter contribution, does not decrease. 

Furthermore, recall that in semiclassical gravity the generalized entropy is defined as the sum of the gravitational entropy and the von Neumann entropy of quantum fields outside the chosen surface. In particular, the von Neumann entropy contains UV divergent local contributions. Focusing on the leading area-law divergence captured by the induced Einstein-Hilbert term in the matter effective action, the sum of the Bekenstein-Hawking term and the leading local divergent part of $S_{\text{vN}}$ can be absorbed into a renormalization of gravitational constant. More precisely, at the level of the leading area term one may write
\begin{equation}
S_{\text{gen}}[\mathcal{C}(v)] \equiv \frac{1}{4G} A[\mathcal{C}(v)] + S_{\text{vN}}
= \frac{1}{4G_{\rm ren}} A[\mathcal{C}(v)] + \cdots \,,
\label{Sgen-leading}
\end{equation}
where the ellipsis denotes additional subleading contributions to the full entropy; in particular, for gauge fields and non-minimally coupled scalars there can be extra terms (e.g.\ contact/edge-mode contributions) that do not contribute to the renormalization of $G$. The corresponding renormalized gravitational constant is
\begin{equation}
    \frac{1}{G_{\rm ren}}=\frac{1}{G}-16\pi c_{2}\Lambda^{d-2}\,.
\label{Gren}
\end{equation}
The inverse of the gravitational constant exhibits a divergence at the 1-loop level, which is canceled by the leading area-law divergence in $S_{\text{vN}}$ \cite{Susskind:1994sm,Bousso:2015mna}. Consequently, the generalized entropy defined in \eqref{Sgen-leading} is finite and physical after renormalization. For stationary black holes, this is the standard generalized entropy satisfying the second law.

This can be carried over to dynamical black holes as follows. Using the interpretation (at the level of the leading local area-law divergence)
\begin{equation}
\tilde{S}_{\text{vN}}=S_{\text{matter}}[\mathcal{T}(v)]
\end{equation}
and $S_{\rm dyn}=S_{\text{grav}}[\mathcal{T}(v)]$ to the first order in perturbations, we define the generalized entropy on the apparent horizon as
\begin{equation}
S_{\text{gen}}[\mathcal{T}(v)] \equiv \frac{1}{4G} A[\mathcal{T}(v)] + \tilde S_{\text{vN}}
= \frac{1}{4G_{\rm ren}} A[\mathcal{T}(v)] + \cdots \,.
\label{SgenApp-leading}
\end{equation}
Then, in the case where the extra contribution in the ellipsis vanishes, such as for minimally coupled scalar field, the GSL \eqref{4.34} can also be expressed as the monotonicity of the generalized entropy evaluated on the (generalized) apparent horizon:
\begin{equation}
\frac{\mathrm{d}}{\mathrm{d}v} \delta S_{\text{gen}}[\mathcal{T}(v)] \geqslant 0\,.
\end{equation}

This formulation naturally extends to $ f(R) $ gravity: each higher order coupling in the Wald entropy $S_{\rm Wald}$ is renormalized by the corresponding higher order term in $S_{\rm vN}$ derived from the matter effective action \cite{Cooperman:2013iqr}. One might therefore expect the generalized entropy, defined using the renormalized couplings evaluated on the (generalized) apparent horizon, to satisfy the GSL. However, the matter effective action in general also contains terms beyond polynomials in the Ricci scalar, such as the terms $ c_{4,2} $ and $ c_{4,3} $ in \eqref{IEmatter}, which would require an analysis in the context of $ f(\text{Riemann}) $ theories.\footnote{The generalized entropy for quadratic gravity has been demonstrated in \cite{Bousso:2015mna}, where the notion of quantum expansion plays the role of the generalized expansion.} We leave this generalization to future work.

\section{Conclusions}
\label{Conclusion}
In this work, we have examined the thermodynamic and entanglement properties of dynamical black holes in both Einstein and $f(R)$ gravity. Working within the framework of GNC, we provided a direct demonstration of the equivalence between the dynamical black hole entropy $S_{\text{dyn}}$ and the Wald entropy $S_{\text{Wald}}$ evaluated on the generalized apparent horizon in $f(R)$ gravity. Combined with the result in Einstein's theory that $S_{\text{dyn}}$ equals the Bekenstein-Hawking entropy of the apparent horizon, this strongly implies that the generalized apparent horizon, rather than the event horizon, serves as the appropriate boundary for a black hole under non-stationary perturbations.

To investigate from the entanglement perspective, we subsequently computed the gravitational entropy using the replica trick, considering both the future event horizon and the apparent horizon as candidate entangling surfaces. Our calculations confirm that while both surfaces yield an area law, only the entropy across the apparent horizon matches the dynamical thermal entropy $S_{\text{dyn}}$ to the first order in perturbations and satisfies the physical process version of the first law. After establishing this result in Einstein gravity, we extended the analysis to $f(R)$ gravity, where we found the same conclusion to hold upon replacing the apparent horizon with its generalized counterpart. This indicates decisively that the apparent horizon serves as the natural entangling surface for dynamical black holes.

Furthermore, we examined the generalized second law for dynamical black holes using the quantum null energy condition. By interpreting the modified von Neumann entropy $\tilde{S}_{\text{vN}}$ introduced in \cite{Hollands:2024vbe} as the matter entanglement entropy across the apparent horizon, we established a clear physical justification for its role. The generalized second law can consequently be reformulated as the statement that the sum of the gravitational entropy and matter field von Neumann entropy across the apparent horizon is non-decreasing. Recognizing this sum as the renormalized Bekenstein-Hawking entropy allows us to identify it with the physical generalized entropy $S_{\text{gen}}$. Thus, the entropy governing the generalized second law for a dynamical black hole is precisely $S_{\text{gen}}$ evaluated at the apparent horizon. 

Looking forward, our results suggest several promising future directions. A natural extension would be to more general gravitational theories, such as $f(\text{Riemann})$ gravity, where additional extrinsic curvature terms could contribute \cite{Dong:2013qoa,Wall:2015raa}. This would require defining generalized expansions that incorporate the full Riemann tensor structure. Given that $S_{\text{dyn}}$ for $f(\text{Riemann})$ gravity receives a correction from $S_{\text{Wall}}$ analogous to the correction from $S_{\text{Wald}}$ in $f(R)$ gravity, it is natural to expect that the gravitational entropy computed from replica method for these theories would correspond to $S_{\text{Wall}}$ evaluated on a generalized notion of apparent horizon. Recently, it has been proposed that entropic marginally outer trapped surfaces (E-MOTSs) could play this role \cite{Furugori:2025pmn}. It would therefore be interesting to study the entropy for higher curvature theories using such surfaces as entangling surfaces.

As with most discussions of dynamical black hole entropy to date, our analysis has been limited to first-order perturbations. Extending the discussion to fully non-perturbative dynamical black holes remains extremely challenging. In such regimes, the notion of an apparent horizon may no longer suffice, and more sophisticated horizon definitions, such as non-expanding horizons, (weakly) isolated horizons, and dynamical horizons \cite{Ashtekar:2000hw,Ashtekar:2000sz,Ashtekar:2001jb,Ashtekar:2003hk,Ashtekar:2004cn} (see also the recent review \cite{Ashtekar:2025wnu}) may be involved. The limitations of analytical methods would also likely require the use of numerical techniques. Such an investigation could reveal how the entanglement prescription of entropy connects to these well-established mathematical frameworks for characterizing evolving black hole boundaries.

Finally, since our analyses of the dynamical black hole entropy from both thermal and entanglement perspectives do not require a vanishing cosmological constant or a specific spacetime asymptotic structure, they remain valid in asymptotically AdS spacetimes. An immediate question arising from our results concerns their implications for the AdS/CFT correspondence in dynamical contexts. The identification of the apparent horizon as the holographic entangling surface suggests that the boundary conformal field theory description of dynamical black hole---including associated phenomena in the AdS bulk like gravitational radiation \cite{Ciambelli:2024kre,Fernandez-Alvarez:2025qqx,Arenas-Henriquez:2025rpt}---should encode information about this locally defined horizon rather than the global event horizon. This perspective naturally motivates a generalization of the Ryu-Takayanagi (RT) and Hubeny-Rangamani-Takayanagi (HRT) formulae \cite{Ryu:2006bv,Hubeny:2007xt} to non-stationary black holes, wherein the bulk minimal or extremal surface should be anchored to the apparent horizon rather than the event horizon.\footnote{The idea of adopting the apparent horizon has been proposed in \cite{Engelhardt:2017aux, Engelhardt:2018kcs}, where the area of the apparent horizon corresponds to the entropy obtained by coarse graining over the information in the black hole interior.} Such a modification could provide new insights into the reconstruction of spacetime geometry from quantum entanglement in non-equilibrium systems.

\section*{Acknowledgements}
We would like to thank Yizhou Lu, Antony Speranza, Qiang Wen and Hongbao Zhang for helpful discussions. This work is supported by a grant from the Research Grants Council of the Hong Kong Special Administrative Region, China (Project No.~AoE/P-404/18).

\appendix
\section{Details of Replica Method Calculation}
\label{Details of entanglement entropy calculation}

In this appendix, we provide some details of the entropy calculation in Subsections \ref{Sec:Future Event Horizon} and \ref{Sec:Apparent Horizon}. We consider a generic Euclidean metric on ${\cal M}_n$ with the following line element:
\begin{align}
\label{metricgeneral}
\td s^{2}&=g_{rr}\beta_{\delta}(r)\td r^{2}+g_{rt}\td r\td t+g_{tt}\td t^{2}+g_{ri}\td r\td x^{i}+g_{ti}\td t\td x^{i}+\gamma_{ij}\td x^{i}\td x^{j}\nn\\
&= (1-a(r,t,x^{i})r)\beta_{\delta}\td r^{2}-2\text{i}a(r,t,x^{i})r^{2}\td r\td t+r^{2}(1+a(r,t,x^{i})r)\td t^{2}+\nn\\&\qquad2c_{i}(r,t,x^{i})\td rdx^{i}+2\I rc_{i}(r,t,x^{i})\td t\td x^{i}+\gamma_{ij}\td x^{i}\td x^{j}\,,
\end{align}
where a regularization function $\beta_{\delta}$ satisfying \eqref{regulatorbeta} is introduced near the conical singularity. In the context of the future event horizon, this metric becomes \eqref{metricFEH} once we recognize $a(r,t,x^{i})=r\alpha(r,t,x^{i}) /4$ and $c_{i}(r,t,x^{i})=r\omega(r,t,x^{i})/2$. In the context of the (generalized) apparent horizon, the above metric gives \eqref{line element of the apparent horizon}, where one should interpret $r,t$ as $\tilde{r},\tilde t$, and take $a(r,t,x^{i})=b'(r,t,x^{i})=\mathcal{O}(\epsilon)+\alpha(r,t,x^{i}) /4$ and $c_{i}(r,t,x^{i})=\mathcal{O}(\epsilon)+r\omega(r,t,x^{i})/2$.

When calculating the Ricci scalar $R$ of the metric \eqref{metricgeneral}, we find that there do not exist terms proportional to $\beta'_{\delta}(r)^{2}$, and thus we can separate $R$ into two parts as follows:
\begin{equation}
\label{Rseparation}
R=R_{0}+\eta\beta'_{\delta}\,,
\end{equation}
where $R_0$ denotes the part without the $r$-derivative of $\beta_{\delta}$, and the function $\eta$ depends on $\beta_\delta$ and the metric functions $a, c_i, \gamma_{ij}$. From its definition \eqref{regulatorbeta}, we can see that $\beta_{\delta}$ is finite when $r<\delta$, and so the integral $\int_{\mathcal{M}_{n},r<\delta}R_{0}$ is bounded by the volume of the region $r<\delta$. Hence, the contribution of this region can be ignored when $\delta\to0$. On the other hand, in the region  $r>\delta$, we have $\beta_{\delta}=1$, and thus the integral $\int_{\mathcal{M}_{n}} R_{0}$ is nothing but $n$ times the integral $\int_{\mathcal{M}} R$ on the original manifold. Therefore, integrating \eqref{Rseparation} over $\mathcal{M}_{n}$ yields
\begin{equation}
\label{intRseparation}
\int _{\mathcal{M}_{n}}R=n\int _{\mathcal{M}}R+\int_{\mathcal{M}_{n}}\eta\beta'_{\delta}\,.
\end{equation}

By definition, $\beta'_{\delta}\neq0$ only when $r<\delta$. Since we will take the limit $\delta\rightarrow0^{+}$, we only have to analyze the behavior of $\eta$ near the horizon at $r\rightarrow0^{+}$. The expression of $\eta$ can be written as
\begin{equation}
\eta(\beta_{\delta},a,c_i,\gamma_{ij})=\frac{\phi(\beta_{\delta},a,c_i,\gamma_{ij})}{\psi(\beta_{\delta},a,c_i,\gamma_{ij})}\,.
\end{equation}
The denominator $\psi$ has the following form:
\begin{equation}
    \psi(\beta_{\delta},a,c_i,\gamma_{ij})=2r(-\beta_{\delta}\det\gamma+\cdots)^{2}\,,
\end{equation}
where the ellipsis contains terms proportional to $c_{i}^{2}\gamma_{jk}$, $c_{i}^{2}\gamma_{jk}\beta_{\delta}$, $rbc_{i}^{2}\gamma_{jk}$ and $rbc_{i}^{2}\gamma_{jk}\beta_{\delta}$. Since $c_{i}(r,t,x^{i})=r\omega_{i}(r,t,x^{i})/2=0$, approaching the horizon at $r\to0^+$ we get
\begin{equation}
\label{psievent}
    \psi(\beta_{\delta},a,c_i,\gamma_{ij})=2r\beta_{\delta}^{2}(\det\gamma)^{2}\,.
\end{equation}
In the case of (generalized) apparent horizon, we have $c_{i}(r,t,x^{i})=\mathcal{O}(\epsilon)+r\omega_{i}(r,t,x^{i})/2=\mathcal{O}(\epsilon)$ as $r\rightarrow0^{+}$ (where $r$ should be interpreted as $\tilde{r}$). Therefore, near the (generalized) apparent horizon, we have
\begin{equation}
\label{psiapparent}
    \psi(\beta_{\delta},a,c_i,\gamma_{ij})=2r(\beta_{\delta}^{2}(\det\gamma)^{2}+\mathcal{O}(\epsilon^{2}))\,.
\end{equation}
The numerator $\phi$ has the following form:
\begin{equation}
\begin{split}
\phi(\beta_{\delta},a,c_i,\gamma_{ij})=2(\det\gamma)^{2}+\cdots\,,
\end{split}
\end{equation}
where the ellipsis contains terms proportional to $c_{i}^{4}\gamma_{jk}^{2}$, $c_{i}^{2}\gamma_{jk}^{3}$, $c_{i}^{3}\partial_{t}c_{j}\gamma_{kl}^{2}$, $c_{i}\partial_{t}c_{j}\gamma_{kl}^{3}$, $c_{i}^{4}\gamma_{jk}\partial_{t}\gamma_{lm}$ and $c_{i}^{2}\gamma_{jk}\partial_{t}\gamma_{lm}$. When $r\rightarrow0^{+}$, we have $\partial_{t}c_{i}(r,t,x^{i})=r\partial_{t}\omega_{i}(r,t,x^{i})=0$. Therefore, it follows from the near horizon behavior of $c_i$ and $\p_tc_i$ that at $r\to0^+$ we get
\begin{equation}
\label{phievent}
\phi(\beta_{\delta},a,c_i,\gamma_{ij})=2(\det\gamma)^{2}\,.
\end{equation}
For the (generalized) apparent horizon, we have $\partial_{t}c_{i}(r,t,x^{i})=\partial_{t}\mathcal{O}(\epsilon)+r\partial_{t}\omega_{i}(r,t,x^{i})=\mathcal{O}(\epsilon)$ when $r\rightarrow0^{+}$ (again, $r$ stands for $\tilde{r}$). Thus, near the (generalized) apparent horizon we have
\begin{equation}
\label{phiapparent}
\phi(\beta_{\delta},a,c_i,\gamma_{ij})=2(\det\gamma)^{2}+\mathcal{O}(\epsilon^{2})\,.
\end{equation}

Finally, we evaluate the integral
\begin{equation}
\int_{\mathcal{M}_{n}}R=\int_{\mathcal{M}_{n}}\td^{d}x\sqrt{\det g}R\,.
\end{equation}
All we have to focus on is the second term in \eqref{intRseparation}, which is localized in the region $r<\delta$. For the future event horizon, we have 
\begin{align}
\int_{\mathcal{M}_{n}}\eta\beta'_\delta&=\int_{\mathcal{M}_{n},r<\delta}\td r\td t\td x^{i}r\sqrt{\beta_{\delta}}\sqrt{\det\gamma}\eta\beta'_\delta\nn\\
&=\int_{\mathcal{M}_{n},r<\delta}\td r\td t\td x^{i}r\sqrt{\beta_{\delta}}\sqrt{\det\gamma}\frac{\beta'_{\delta}}{r\beta_{\delta}^{2}}\nn\\
\label{A.14}
&=-2\int_{\mathcal{M}_{n},r<\delta}\td r\td t\td x^{i}\sqrt{\det}\partial_{r}(\beta_{\delta}^{-\frac{1}{2}})\,,
\end{align}
where we used \eqref{detgapprox} in the first equality, and used \eqref{psievent} and \eqref{phievent} in the second equality. The smooth function $\beta_{\delta}^{-\frac{1}{2}}$ can be constructed by means of the trigonometric function such that it satisfies the requirement of $\beta_{\delta}$ \eqref{regulatorbeta}:
\begin{equation}
    \beta_{\delta}^{-\frac{1}{2}} =\begin{cases}\frac{1}{n} +(1-\frac{1}{n} )\text{sin}^{2} (\frac{r}{\delta} \frac{\pi}{2} )&r\leqslant \delta\,,\\ 1&r>\delta\,.\end{cases}
\end{equation} 
In the limit of $\delta\to0^{+}$, it can be shown that
\begin{equation}
    \beta_{\delta}^{-\frac{1}{2}}=\frac{1}{n}+\Big(1-\frac{1}{n}\Big)\theta(r-0^{+})\,.
    \label{A.15}
\end{equation}
This result does not depend on the specific choices of $\beta_{\delta}$.
Plugging \eqref{A.15} into \eqref{A.14} we find that
\begin{align}
\int_{\mathcal{M}_{n}}\eta\beta'_\delta=&-2\int_{\mathcal{M}_{n}, r<\delta}\td r\td t\td x^{i}\sqrt{\det\gamma}\partial_{r}(\beta_{\delta}^{-\frac{1}{2}})\\
    &=\frac{2(1-n)}{n}\int_{\mathcal{M}_{n}, r<\delta}\td r\td t\td x^{i}\sqrt{\det\gamma}\delta(r-0^{+})\\
    &=4\pi(1-n)\int_{\mathcal{C}}1=4\pi(1-n){\cal A}[C]\,.
\end{align}
This gives the result shown in \eqref{results in future horizon}.

For the (generalized) apparent horizon, the discussion is similar. Using \eqref{psiapparent} and \eqref{phiapparent}, one finds (notice again that $r,t$ stands for $\tilde r,\tilde t$)
\begin{align}
\int_{\mathcal{M}_{n}}\eta\beta'_\delta&=\int_{\mathcal{M}_{n},r<\delta}\td r\td t\td x^{i}r\sqrt{\beta_{\delta}}\sqrt{\det\gamma}\frac{\beta'_{\delta}}{r\beta_{\delta}^{2}}+\mathcal{O}(\epsilon^{2})\nn\\
&=-2\int_{\mathcal{M}_{n},r<\delta}\td r\td t\td x^{i}\sqrt{\det\gamma}\partial_{r}(\beta_{\delta}^{-\frac{1}{2}})+\mathcal{O}(\epsilon^{2})\nn\\
&=\frac{2(1-n)}{n}\int_{\mathcal{M}_{n},r<\delta}\td r\td t\td x^{i}\sqrt{\det \gamma}\delta(r-0^{+})+\mathcal{O}(\epsilon^{2})\nn\\
&=4\pi(1-n)\int_{\mathcal{T}}1+\mathcal{O}(\epsilon^{2})=4\pi(1-n)A[\mathcal{T}]+\mathcal{O}(\epsilon^{2})\,.
\end{align}
This gives the result shown in \eqref{results in apparent horizon}.

\section{Killing Symmetry and Boost Weight}
\label{Rescaling Gauge Freedom and Boost Weight}

In a stationary background geometry, the metric in \eqref{2.7} depends on $u,v$ only through the product $uv$ in the affine GNC system. Such a metric is invariant under the rigid rescaling $u \to qu$, $v \to v/q$. This transformation corresponds to the isometry generated by the horizon Killing vector $\xi^a$ defined in \eqref{Killing}, which acts as a standard Lorentz boost. For any quantity $Z$ that transforms as $Z \to q^w Z$ under $u \to qu$, $v \to v/q$, we say that it has a \emph{Killing weight} or \emph{boost weight} $w$ \cite{Wall:2015raa, Wall:2024lbd, Hollands:2022fkn, Bhattacharyya:2021jhr}. For instance, given a tensor of type $(k,l)$ with components $T^{\mu_1\cdots \mu_k}{}_{\nu_1\cdots \nu_l}$, its boost weight $w$ can be read off as follows:
\begin{align}
w &= \text{\#} \text{ lower indices of }v-\text{\#} \text{ upper indices of }v\nn \\
&\quad\,+ \text{\#} \text{ upper indices of }u-\text{\#} \text{ lower indices of }u\,. 
\end{align}
In the GNC system, the Lie derivative of a quantity $Z$ with boost weight $w$ along the Killing vector field $\xi^a$ is (see Appendix~B of \cite{Bhattacharyya:2021jhr} for a proof)
\begin{equation}
\mathcal{L}_\xi Z = \kappa(v\partial_v - u\partial_u + w)Z\,. 
\end{equation}
Suppose $Z$ satisfies the stationarity condition $\mathcal{L}_\xi Z=0$. Then, the general solution to
\begin{equation}
(v\partial_v - u\partial_u + w)Z=0
\end{equation}
can be found to be
\begin{equation}
Z(u,v,x^i)=\sum_{n\geqslant -w}u^{n+w}v^n F^Z_n(x^i)\,.
\end{equation}
On the horizon $u= 0$, only the term with $n=-w$ survives, and we have
\begin{equation}
\label{B3}
Z(0,v,x^i)=v^{-w} F^Z_{-w}(x^i)\,.
\end{equation}
Note that for $w>0$, one needs to require $F^Z_{-w}(x^i)=0$ to make the solution regular at the bifurcation surface $u=v=0$, which leads to $Z(0,v,x^i)=0$.

We now apply the above analysis to the ingoing null generalized expansion $\Theta_{u}$ in $f(R)$ gravity generated by $l^a = (\partial_u)^a$. (Note that the standard expansion $\theta_{u}$ can be viewed as a special case of $\Theta_{u}$ with $f(R)=R$.) Recall that from \eqref{definition of generalized expansion}, the generalized expansion $\Theta_{u}$ is defined as
\begin{equation}
    \Theta_{u}=\partial_{u}\log(\sqrt{\det\gamma}f'(R))\,.
\end{equation}
It is easy to check that $\det\gamma$ is invariant under the boost transformation, and $f'(R)$ is obviously invariant as a Lorentz scalar. Therefore, under the transformation $u \to qu$, $v \to v/q$, the generalized expansion transforms as $\Theta_{u} \to q^{-1} \Theta_{u}$, i.e., it possesses weight $-1$. Then, it follows from \eqref{B3} that on the future event horizon we have
\begin{equation}
    \Theta_{u}(v,0,x^i)=vF^{\Theta}_{1}(x^{i})\,.
\end{equation}
For the standard expansion $\theta_u$ (i.e., $f(R)=R$), the function $F^\theta_{1}(x^i)$ is guaranteed to be negative since null geodesics are focusing in the $u$-direction.


\providecommand{\href}[2]{#2}\begingroup\raggedright\endgroup

\end{document}